\documentclass[12pt]{iopart}

\newcommand{\bea}{\begin{eqnarray}}
\newcommand{\eea}{\end{eqnarray}}
\usepackage{amssymb}
\usepackage{bm}
\usepackage{axodraw}
\usepackage[section]{placeins}

\begin{document}

\title{Gauge Threshold Corrections in Warped Geometry}

\author{Kiwoon Choi$^1$, Ian-Woo Kim$^2$ and Chang Sub Shin$^1$}

\address{$^1$Department of Physics, KAIST, Daejeon 305-701, Korea}

\address{$^2$Department of Physics, University of Wisconsin, Madison, WI 53706, USA}

\ead{kchoi@kaist.ac.kr$,$ ikim@physics.wisc.edu $\rm and$ csshin@muon.kaist.ac.kr}


\begin{abstract}
We discuss the Kaluza-Klein threshold correction
to low energy gauge couplings in theories with warped extra-dimension,
which might be crucial for the gauge coupling unification when the
warping is sizable. Explicit expressions of one-loop thresholds
are derived for generic 5D gauge theory on a slice of $AdS_5$, where
some of the bulk gauge symmetries are broken by orbifold boundary
conditions and/or by bulk Higgs vacuum values. Effects of the mass
mixing between the bulk  fields with different orbifold parities are
included as such mixing is required in some class of realistic
warped unification models.

\end{abstract}

\maketitle

\section{Introduction}

In theories with unified gauge symmetry at high energy scale,
threshold corrections due to heavy  particles often affect the
predicted low energy gauge couplings
significantly~\cite{Hall:1980kf}. Since symmetry breaking leads to a
mass splitting between the particles in an irreducible
representation of the unified gauge symmetry, the low energy gage
couplings generically acquire  non-universal quantum corrections
when the heavy particles are integrated out. In four-dimensional
(4D) theories, the resulting differences between low energy gauge
couplings are proportional to the logarithm of the mass ratios.
Therefore, those threshold effects can be particularly important
when the mass splitting occurs over a wide range of energy scales
and/or for many numbers of massive particles.



Such situation can be realized 
in higher dimensional gauge theories (including string theories), in
which there exist generically an infinite tower of gauge-charged
Kaluza-Klein (KK) states. Higher dimensional gauge theories can
employ a novel class of symmetry breaking mechanisms such as the one
by boundary condition~\cite{Scherk:1978ta} or by the vacuum
expectation value (VEV) of the extra-dimensional component of gauge
field~\cite{Hosotani:1983xw}. Such mechanisms might successfully
address various naturalness problems of grand unified theories
(GUTs)~\cite{Witten:1985xc} and/or explain the origin of the Higgs
field~\cite{Hatanaka:1998yp}. In higher dimensional theories with
broken gauge symmetry, the whole KK tower of higher dimensional
fields  are splitted. This splitting can yield a large threshold
correction because of the infinite number of KK modes and also a
large scale difference between the lowest KK mass and the cutoff
scale of the theory~\cite{Hall:2001pg}.

On the other hand, calculation of the KK thresholds  requires a
careful treatment of the associated UV divergences. Summing up the
logarithmic contribution from each KK mode, it is expected that
power-law-divergent contributions appear~\cite{Dienes:1998vh}. In
field theory, all the ultraviolet (UV) divergences must be absorbed
into local counterterms that are consistent with the defining
symmetry of the theory.
 In models with unified gauge symmetry in bulk spacetime,
those power-law divergences are universal and can be absorbed into a
renormalization of the unified higher dimensional gauge coupling at
the cutoff scale $\Lambda$.
However, if the unified gauge symmetry is broken  by a boundary
condition at the orbifold fixed point,
there can be non-universal logarithmically-divergent counterterms
localized at the fixed point. Those logarithmic divergences are
associated with the renormalization group (RG) runnings of the
fixed-point gauge coupling constants~\cite{Georgi:2000ks}, which
lead to a controllable consequence in the predicted low-energy gauge
couplings as in the case of  conventional 4D
GUTs~\cite{Hall:2001pg}. After identifying the UV-divergent pieces
of the KK threshold corrections, the finite calculable parts are
unambiguously defined.\footnote{In string theory, the full threshold
corrections including stringy thresholds are finite with the cutoff
scale $\Lambda$ replaced by the string scale. For an early
discussion of threshold corrections in compactified string theory,
see for instance~\cite{Nilles:1986cy,Kaplunovsky:1987rp}.}
 In general,
those finite corrections heavily depend on the parameters of the
model, including the symmetry breaking vacuum expectation values
(VEVs) and the masses of higher dimensional fields, as well as on
the structure of the background spacetime geometry.

It has been of particular interest to study quantum corrections in
warped geometry. Warped extra-dimension might be responsible for the
weak scale to the Planck scale hierarchy~\cite{Randall:1999ee}, or
the supersymmetry breaking scale to the Planck scale
hierarchy~\cite{Luty:2000ec,Kachru:2003aw}, or even the Yukawa
coupling hierarchies~\cite{Huber:2000ie}. There also have been
studies on higher dimensional GUTs in warped geometry, showing quite
distinct features arising from  the
warping~\cite{Pomarol:2000hp,Goldberger:2002pc,Carena:2003fx,Agashe:2003zs,Agashe:2004ci}.
In warped models, gauge threshold corrections might be crucial for a
successful unification
when the lowest KK scale $m_{KK}$ is hierarchically lower than the
conventional unification scale $M_{GUT}\sim 2\times 10^{16}$ GeV.
A series of studies on quantum corrections in anti-de Sitter space
(AdS) show that KK threshold corrections in warped gauged theory is
enhanced by the large logarithmic factor  $\ln
(e^\Omega)$~\cite{Pomarol:2000hp,Randall:2001gb,Choi:2002wx,Falkowski:2002cm,Choi:2002ps},
where $e^\Omega$ is an exponentially small warp factor. Explanation
of this logarithmic factor has been attempted in various contexts,
including those based on
the AdS/CFT correspondence which states that a 5D theory on a slice
of AdS$_5$ can be regarded as a 4D conformal field theory (CFT) with
conformal symmetry spontaneously broken at $m_{KK}$
~\cite{Maldacena:1997re,ArkaniHamed:2000ds}.

In \cite{Choi:2002ps}, a novel method to compute  1-loop gauge
couplings in higher dimensional gauge theory with warped extra
dimension has been discussed, and explicit analytic expressions of
the KK thresholds in 5D  theory on a slice of AdS$_5$ have been
derived for the case that some part of bulk gauge symmetries are
broken by orbifold boundary condition with no mass mixing between
bulk fields with different orbifold parities. In this paper, we wish
to extend the analysis of \cite{Choi:2002ps} to  more general case
including  the possibility of symmetry breaking by bulk scalar VEVs
and also of non-zero mass mixing among bulk fields with different
orbifold parities.
 Our results  then cover most of the warped GUT models discussed so far in the
literatures.

The organization of this paper is as follows. In the next section,
we first discuss some features of KK thresholds which are relevant
for our later discussion, and then examine a simple example of 5D
scalar threshold to illustrate our computation method. In Section 3,
we consider generic 5D gauge theory defined on a slice of AdS$_5$,
and derive analytic expression of 1-loop KK thresholds induced by 5D
gauge and matter fields when some part of the bulk gauge symmetries
are broken by orbifold boundary conditions and/or by bulk Higgs
vacuum values. To be general, we also include the effects of mass
mixing between the bulk fields with different orbifold parities. In
Section 4, we give a conclusion. We provide in Appendix A a detailed
discussion of the $N$-function whose zeros correspond to the KK
spectrum, and a discussion of boundary matter fields in Appendix B.

\section{Some generic features of Kaluza-Klein threshold corrections}

The 5D gauge theory in consideration can be defined as a Wilsonian
effective field theory with the action
  \bea \label{action} S_W&=&-\int d^5x \sqrt{-G}\,
\frac{1}{4}\left(\frac{1}{g_{5a}^2}+\frac{\kappa_a}{4\pi^2}\frac{\delta(y)}{\sqrt{G_{55}}}+\frac{\kappa_a^\prime}{4\pi^2}
\frac{\delta(y-\pi)}{\sqrt{G_{55}}} \right)F^{aMN}F^a_{MN}\nonumber
\\
&+& S_{\rm gauge-fixing}+S_{\rm ghost}+S_{\rm matter}, \eea
 where $S_{\rm gauge-fixing}$ and $S_{\rm ghost}$ are the
 gauge-fixing term and the associated ghost action, respectively,
 $S_{\rm matter}$ is the model-dependent action of 5D
scalar and fermion
  matter fields, and
 the 5D spacetime metric $G_{MN}$ is assumed to take a generic 4D Poincare-invariant
 form:
 \bea ds^2=G_{MN}dx^Mdx^N =e^{2\Omega(y)}\eta_{\mu\nu}dx^\mu dx^\nu + R^2 dy^2
\quad (0\leq y \leq \pi), \nonumber \eea where  $\pi R$ is the
proper distance of the interval, $\eta_{\mu\nu}$ corresponds to the
4D graviton zero mode which is used by low energy observer to
measure the external 4D momentum $p^\mu$ as well as the KK mass
spectrum, and  we are using the warp factor convention:
$e^{\Omega(y=0)}=1$ and $e^{\Omega(y=\pi)}\leq 1$.
 Here we do not include any boundary matter
field separately as it can be considered as the localized limit of
bulk matter field, which is achieved by taking some mass parameters
to the cutoff scale. (For a discussion of this point, see  Appendix
B.) Note that the range of the 5-th dimension is taken as $0\leq
y\leq \pi$ with the convention: $\int_0^\pi dy \delta(y)=\int_0^\pi
dy\delta(y-\pi)=1/2$.

 In order
for the theory to be well-defined, one also needs to specify  the UV
cutoff scheme along with the Wilsonian action.
Then all the Wilsonian couplings in $S_W$ depend implicitly on the
associated cutoff scheme $\Lambda$, and this $\Lambda$-dependence of
Wilsonian couplings should cancel the $\Lambda$-dependence of
regulated quantum corrections, rendering all the observable quantities
to be independent of $\Lambda$.

The quantity  of our concern is the low energy one-particle-irreducible (1PI)
gauge couplings of
 4D gauge boson zero modes.  It can be obtained by evaluating \bea
e^{i\Gamma [\Phi_{cl}]} =\int \left[{\cal D}\Phi_{qu}\right]\,
e^{iS_W[\Phi_{cl}+\Phi_{qu}]} \eea where $\Phi_{cl}$ denotes
background field configuration which includes the 4D gauge boson
zero modes  $A_{\mu}^{a(0)}$ as well as the vacuum values of scalar
fields, and $\Phi_{qu}$ stands for quantum fluctuations of the  5D
gauge, matter and ghost fields in the model.
 The
resulting 1PI gauge coupling $g_a^2(p)$ of  $A_{\mu}^{a(0)}(p)$
carrying an external 4D momentum $p^\mu$ is given by \bea \frac{
(-p^2\eta^{\mu\nu}+p^\mu p^\nu)}{g_a^2(p)} \equiv
\left.\frac{\delta^2\Gamma}{\delta A_\mu^{a(0)}(p)\delta
A_\nu^{a(0)}(-p)}\right|_{A_\mu^{a(0)}=0}. \eea As the gauge boson
zero modes have a constant wavefunction over the 5th dimension, the
4D gauge couplings at tree level are simply given by \bea
\label{tree} \left(\frac{1}{g_a^2}\right)_{\rm tree}=\frac{\pi
R}{g_{5a}^2}+\frac{1}{8\pi^2}\left(\kappa_a+\kappa_a^\prime\right).\eea
To compute quantum corrections, one needs to introduce a suitable
regularization scheme which might involve a set of regulator masses
collectively denoted by $\Lambda$\footnote{At this stage, we assume
a mass-dependent cutoff scheme introducing an appropriate set of
Pauli-Villars regulating fields and/or  higher derivative regulating
terms, although  eventually we will use a mass-independent
dimensional regularization which is particularly convenient for the
computation of gauge boson loops.}. One also needs to deal with  a
summation over the  KK modes whose mass eigenvalues $\{m_n\}$ depend
on various model parameters  that will be collectively denoted by
$\lambda$, for instance  the bulk or boundary masses of the matter
and gauge fields as well as the AdS vacuum energy density that would
determine the warp factor. Note that the 4D momentum $p^\mu$ of the
gauge boson zero modes  and  the KK mass eigenvalues $\{m_n\}$ are
defined in the 4D metric frame of the graviton zero mode
$\eta_{\mu\nu}$, while $\Lambda$, $\lambda$ and $1/R$ are the 5D
mass parameters invariant under the 5D general coordinate
transformation.

 Schematically, one-loop
correction to the 4D 1PI gauge coupling is given by \bea
\frac{1}{8\pi^2}\Delta_a(p,\Lambda,R,\lambda)
=\sum_{\Phi_0,\Phi_n}\int \frac{d^4l}{(2\pi)^4} \,
f_a(p,l,m_n(R,\lambda)),\eea where $\Phi_0$ denotes the light zero
modes with a mass $m_0\ll p$, while $\Phi_n$ stands for  the massive
KK modes with $m_n\gg p$.
In the limit $p\ll m_{KK}$ and $\Lambda\gg \lambda$, where $m_{KK}$
is the lowest KK mass, the above 1-loop correction takes the
form~\cite{Hall:2001pg,Choi:2002wx} \bea \frac{1}{8\pi^2}\Delta_a
&=& \frac{\gamma_a}{24\pi^3}\Lambda\pi R+\frac{1}{8\pi^2}\left[
\tilde{b}_a\ln(\Lambda\pi R)-b_a\ln(p\pi
R)+ \tilde{\Delta}_a(R, \lambda)\right] \nonumber \\
&+& {\cal O}\left(\frac{p^2}{m_{KK}^2}\right)+{\cal
O}\left(\frac{\lambda}{\Lambda},\frac{1}{\Lambda R}\right). \eea
Then the low energy 1PI couplings are given by
 \bea
\frac{1}{g_a^2(p)}&=& \left(\frac{1}{g_a^2}\right)_{\rm
tree}+\frac{1}{8\pi^2}\Delta_a\nonumber
 \\
&=& \frac{\pi R}{\hat{g}_{5a}^2} +\frac{1}{8\pi^2}\hat{\kappa}_a(\ln
p, \lambda, R)+ {\cal O}\left(\frac{p^2}{m_{KK}^2}\right)+{\cal
O}\left(\frac{\lambda}{\Lambda},\frac{1}{\Lambda R}\right), \eea
where \bea \label{def} \frac{1}{\hat{g}_{5a}^2}&=&
\frac{1}{g_{5a}^2(\Lambda)}+\frac{\gamma_a}{24\pi^3}\Lambda
\nonumber\\
\hat{\kappa}_a
&=&\kappa_a(\Lambda)+\kappa^\prime_a(\Lambda)+\tilde{b}_a\ln(\Lambda
\pi R)-b_a\ln\left({p\pi R}\right) +\tilde\Delta_a(R,\lambda ). \eea
The above expression of 4D 1PI coupling is valid only for $p <
m_{KK}$. However, it still provides a well-defined matching  between
the observable low energy gauge couplings and the fundamental
parameters in the 5D action defined at the cutoff scale $\Lambda\gg
m_{KK}$. Note that the Wilsonian couplings $g_{5a}^2,
\kappa_a,\kappa_a^\prime$ depend on $\Lambda$ in such a way to make
$\hat{g}_{5a}^2$ and $\hat{\kappa}_a$ to be independent of
$\Lambda$.


The linearly divergent piece  in (\ref{def}) originates from the KK
modes around the cutoff scale $\Lambda$, and therefore its
coefficient $\gamma_a$ severely depends on the employed cutoff
scheme. For instance, in a mass-dependent cutoff scheme introducing
an appropriate set of Pauli-Villars (PV) regulating fields and/or
higher derivative regulating terms,  each $\gamma_a$ has a nonzero
value depending on the detailed structure of the regulator masses
and the regulator coefficients, while it vanishes in a
mass-independent cutoff scheme such as dimensional
regularization~\cite{GrootNibbelink:2001bx}\footnote{A novel
extension of dimensional regularization for higher dimensional gauge
theory has been suggested also in \cite{Bauman:2007rt}. }.
 Note that this
does not affect the calculable prediction of the theory, which is
determined by the scheme-independent combination $1/\hat{g}_{5a}^2$.

Unlike the coefficient of power-law divergence, the coefficients of
$\ln p$ and $\ln \Lambda$ are unambiguously determined by the
physics below $\Lambda$~\cite{Hall:2001pg,Choi:2002wx}.
As $\ln p$ originates from the light zero modes with $m_0\ll p$, one
immediately finds  \bea \label{ba} b_a=
\frac{1}{6}\sum_{\varphi^{(0)}} {\rm
Tr}(T_a^2(\varphi^{(0)}))+\frac{2}{3}\sum_{\psi^{(0)}}{\rm
Tr}(T_a^2(\psi^{(0)}))-\frac{11}{3}\sum_{A^{(0)}_{\mu}}{\rm Tr}(
T^2_a(A^{(0)}_\mu)),\eea where $\varphi^{(0)}$, $\psi^{(0)}$ and
$A^{(0)}_{\mu}$ denote the 4D real scalar, 4D chiral fermion and 4D
real vector boson zero modes which originate from 5D matter and
gauge fields, and
 $T_a(\Phi)$ is the generator of the unbroken
gauge transformation of $\Phi$.
 Note that $\varphi^{(0)}$ can originate from a 5D
vector field.

 The logarithmic   divergence appears
 because of  the orbifold fixed points.
This implies that $\tilde{b}_a$ are determined just by the orbifold
boundary condition of 5D fields if there is no 4D matter field
confined  at the fixed point. The logarithmic divergence generically
takes the form \bea \label{log div}  -\int d^5 x\sqrt{-G}\,
\frac{\ln\Lambda}{16\pi^2}\left(
\frac{\lambda_{a0}\delta(y)}{\sqrt{G_{55}}}+\frac{\lambda_{a\pi}\delta(y-\pi)}{\sqrt{G_{55}}}\right)
F^a_{\mu\nu}F^{a\mu\nu}, \eea and the coefficients $\lambda_{a0}$
and $\lambda_{a\pi}$ are independent of the smooth geometry of the
underlying spacetime.
 It is then straightforward to
determine  $\lambda_{a0}$ and $\lambda_{a\pi}$ in the flat orbifold
limit, which yields~\cite{Hall:2001pg,Choi:2002wx}
 \bea \lambda_{a0} &=& \sum_{zz'}
\frac{z}{24}\Big({\rm Tr}(T_a^2(\phi_{zz'})) -23{\rm
Tr}(T_a^2(A^M_{zz'}))\Big),\nonumber \\
\lambda_{a\pi} &=& \sum_{zz'} \frac{z^\prime}{24}\Big({\rm
Tr}(T_a^2(\phi_{zz'})) -23{\rm Tr}(T_a^2(A^M_{zz'}))\Big), \eea and
thus \bea \label{logd} \tilde{b}_a= \lambda_{a0}+\lambda_{a\pi}=
\sum_{zz'} \frac{(z+z')}{24} \Big({\rm Tr}(T_a^2(\phi_{zz'}))
-23{\rm Tr}(T_a^2(A^M_{zz'}))\Big), \eea where $\phi_{zz^\prime}$
and $A^M_{zz^\prime}$ ($z,z^\prime=\pm 1$) denote 5D real scalar and
vector fields with the orbifold boundary condition: \bea
\label{parity} &&\phi_{zz^\prime}(-y) = z\phi_{zz^\prime}(y), \quad
\phi_{zz^\prime}(-y+\pi)=z^\prime \phi_{zz^\prime}(y+\pi),
\nonumber \\
&& A^M_{zz^\prime}(-y)=z\epsilon_M A^M_{zz^\prime}(y),\quad
A^M_{zz^\prime}(-y+\pi)=z^\prime \epsilon_M A^M_{zz^\prime}(y+\pi),
\eea where $\epsilon_\mu=1$ and $\epsilon_5=-1$.

The last part of 1-loop correction, i.e.
$\tilde\Delta_a(R,\lambda)$, is highly model-dependent as it
generically depends on various parameters of the underlying 5D
theory, e.g. the curvature of background geometry, matter and gauge
field masses in the bulk and at the boundaries, and also on the
orbifold boundary conditions of 5D fields. Note that all of these
features affect the KK mass spectrum, and thus the KK thresholds. In
many cases, it can be an important part of quantum correction, even
a dominant part in warped case. The aim of this paper is to provide
an explicit expression of $\tilde\Delta_a$ as a function of the
fundamental parameters in  5D theory in a general context as much
as possible.



Let us now consider a specific  example of 5D scalar threshold to
see some of the features discussed above. We start with the case of
a massless 5D {\it complex} scalar field $\phi_{zz^\prime}$ in the
flat spacetime background:
 \bea S_{\rm matter}&=&-\int d^5x \sqrt{-G}
\sum_{z,z^\prime}G^{MN}D_M\phi^{\dagger}_{zz^\prime}
D_N\phi_{zz^\prime},\eea where \bea
ds^2=G_{MN}dx^Mdx^N=\eta_{\mu\nu}dx^\mu dx^\nu+R^2dy^2.\nonumber
\eea In this case, one can easily find an explicit form of the KK
spectra: \bea m_n(\phi_{zz'}) = \left\lbrace
\begin{array}{ll} \frac{n}{R} & \mbox{for} \quad \phi_{++} \\
\frac{2n+1}{2R} & \mbox{for} \quad \phi_{+-}\quad \mbox{and} \quad \phi_{-+}
\\ \frac{n+1}{R} & \mbox{for} \quad \phi_{--} .\end{array}
\right., \eea where $n$ is a non-negative integer.  The
corresponding 1-loop correction can be obtained using a simple
momentum cutoff: \bea \label{scalar_threshold}
\frac{1}{8\pi^2}(p^2\eta^{\mu\nu}-p^\mu p^\nu)
\Delta_a(\phi_{zz^\prime}) = \sum_{z,z^\prime}\sum_{n=0}^{\Lambda R}
{\rm Tr}(T_a^2(\phi_{zz^\prime})) \int \frac{d^4l}{(2\pi)^4}
f^{\mu\nu},\eea where \bea f^{\mu\nu}=
\frac{2\eta^{\mu\nu}((p+l)^2+m_n^2(\phi_{zz^\prime}))-(p+2l)^\mu(p+2l)^\nu}{i((p+l)^2+m_n^2(\phi_{zz^\prime}))(l^2+
m_n^2(\phi_{zz^\prime}))}, \nonumber \eea   which gives (in the
limit $p\ll m_{KK}=1/R$) \bea \label{massless} \Delta_a&=&
\frac{1}{3}{\rm
Tr}(T_a^2(\phi_{++}))\ln\left(\frac{\Lambda}{p}\right)\nonumber
\\
&+&\frac{1}{3}\Big[{\rm Tr}(T_a^2(\phi_{++}))+ {\rm
Tr}(T_a^2(\phi_{--}))\Big]\sum_{n=1}^{\Lambda  R}
\ln\left(\frac{\Lambda R}{n}\right)\nonumber
\\
&+&\frac{1}{3}\Big[{\rm Tr}(T_a^2(\phi_{+-}))+{\rm
Tr}(T_a^2(\phi_{-+}))\Big]\sum_{n=1}^{\Lambda
R}\ln\left(\frac{2\Lambda R}{2n-1}\right)+{\cal O}(1)
\nonumber \\
&=& \frac{1}{3}\Big[ {\rm Tr}(T_a^2(\phi_{++}))+{\rm
Tr}(T_a^2(\phi_{--}))+{\rm Tr}(T_a^2(\phi_{+-}))+{\rm
Tr}(T_a^2(\phi_{-+})) \Big]\Lambda
R\nonumber \\
&+& \frac{1}{6}\Big[{\rm Tr}(T_a^2(\phi_{++}))-{\rm
Tr}(T_a^2(\phi_{--}))\Big]\ln (\Lambda \pi R) \nonumber \\
&-&\frac{1}{3}{\rm Tr}(T_a^2(\phi_{++}))\ln (p\pi R)+ {\cal O}(1).
\eea Obviously, in case with a unified gauge symmetry in bulk
spacetime, the coefficients of linear divergence, i.e.
$\sum_{z,z^\prime}{\rm Tr}(T_a^2(\phi_{zz^\prime}))$, are universal.
Also the above result gives \bea \tilde{b}_a=\frac{1}{6}\Big[{\rm
Tr}(T_a^2(\phi_{++}))-{\rm Tr}(T_a^2(\phi_{--}))\Big],\nonumber\eea
which confirms the result of (\ref{logd}). Note that
$\phi_{zz^\prime}$ here are complex scalar fields, while
$\phi_{zz^\prime}$ in (\ref{logd}) are real scalar fields.


One can generalize the above result by introducing a nonzero bulk
mass. To see the effect of bulk mass, let us consider  $\phi_{++}$
with a 5D mass $M_S\gg p$ in the flat spacetime background. It is
still straightforward to find the explicit form of KK spectrum:\bea
m_n = \sqrt{M_S^2 + \frac{n^2}{R^2}}.\eea
In this case,  there is no light mode since $M_S\gg p$, and
therefore $b_a=0$. Again the 1-loop threshold can be computed with a
simple momentum cut off: \bea \Delta_a(\phi_{++}) &=& \frac{1}{6}
{\rm Tr}(T_a^2(\phi_{++})) \sum_{n=0}^{\Lambda
R}\ln\left(\frac{\Lambda^2R^2}{M_S^2R^2 + n^2}\right) +
{\cal O}(1) \nonumber \\
&=& \frac{1}{6} {\rm Tr}(T_a^2(\phi_{++}))\left[\,2\Lambda R +
\left(\ln\frac{\Lambda}{M_S}- \ln \sinh(M_S\pi R)\right) + {\cal
O}(1)\,\right]. \nonumber \eea

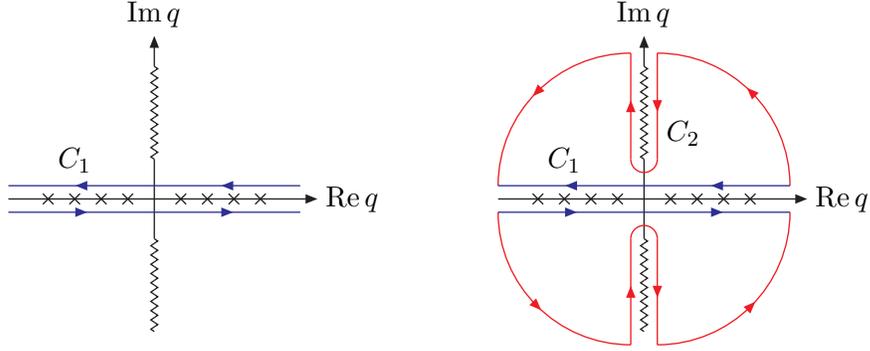
\begin{figure}
\bea\hskip 1cm
\begin{picture}(150,120)
\SetScale{1} \LongArrow(-5,50)(110,50) \Line(50,35)(50,65)
\LongArrow(50,100)(50,110) \ZigZag(50,0)(50,35){1.3}{10}
\ZigZag(50,65)(50,100){1.3}{10} \Line(8,48)(12,52)
\Line(12,48)(8,52) \Line(18,48)(22,52) \Line(22,48)(18,52)
\Line(28,48)(32,52) \Line(32,48)(28,52) \Line(38,48)(42,52)
\Line(42,48)(38,52) \Line(58,48)(62,52) \Line(62,48)(58,52)
\Line(68,48)(72,52) \Line(72,48)(68,52) \Line(78,48)(82,52)
\Line(82,48)(78,52) \Line(88,48)(92,52) \Line(92,48)(88,52)
\Text(50,120)[]{\small Im$\, q$} \Text(125,50)[]{\small Re$\, q$}
\Text(20,65)[]{\small $C_1$} \SetColor{Blue}
\ArrowLine(105,55)(50,55) \ArrowLine(50,55)(-5,55)
\ArrowLine(50,45)(105,45) \ArrowLine(-5,45)(50,45)
\end{picture}
&& \quad\quad\quad
\begin{picture}(150,120)
\SetScale{1} \LongArrow(-5,50)(110,50) \Line(50,35)(50,65)
\LongArrow(50,100)(50,110) \ZigZag(50,0)(50,35){1.3}{10}
\ZigZag(50,65)(50,100){1.3}{10} \Text(50,120)[]{\small Im$\, q$}
\Text(125,50)[]{\small Re$\, q$} \Text(20,65)[]{\small $C_1$}
\Text(65,75)[]{\small $C_2$} \Line(8,48)(12,52)  \Line(12,48)(8,52)
\Line(18,48)(22,52) \Line(22,48)(18,52) \Line(28,48)(32,52)
\Line(32,48)(28,52) \Line(38,48)(42,52) \Line(42,48)(38,52)
\Line(58,48)(62,52) \Line(62,48)(58,52) \Line(68,48)(72,52)
\Line(72,48)(68,52) \Line(78,48)(82,52) \Line(82,48)(78,52)
\Line(88,48)(92,52) \Line(92,48)(88,52) \SetColor{Red}
\ArrowArc(55,55)(50,0,90) \ArrowArc(45,55)(50,90,180)
\ArrowLine(55,105)(55,65) \ArrowLine(45,65)(45,105)
\CArc(50,65)(5,180,360) \ArrowArc(45,45)(50,180,270)
\ArrowArc(55,45)(50,270,360) \ArrowLine(45,-5)(45,35)
\ArrowLine(55,35)(55,-5) \CArc(50,35)(5,0,180) \SetColor{Blue}
\ArrowLine(105,55)(50,55) \ArrowLine(50,55)(-5,55)
\ArrowLine(50,45)(105,45) \ArrowLine(-5,45)(50,45)
\end{picture} \nonumber
\eea \caption{\label{contour} \small Integration contours on the
$q$-plane. Crosses along the real axis represent the KK masses
$\{\pm m_n\}$ which correspond to the poles of $P(q)$. The branch
cut along the imaginary axis arises from ${\cal G}_a(p,q)$, and the
contour $C_1$ can be deformed to the contour $C_2$ for the
integration involving $P_{\rm finite}(q)$.}
\end{figure}

For warped spacetime background,
the KK spectrum takes a more complicate form, and its explicit form
is usually not available. Furthermore, as the 4D loop momentum
$l^\mu$ and the KK spectrum $\{m_n\}$ are defined in the metric
frame of 4D graviton zero mode, the cutoff scales for $l^\mu$ and
$\{m_n\}$ depend on the position in warped extra-dimension. One can
avoid these difficulties using the Pole function method with
dimensional
regularization~\cite{GrootNibbelink:2001bx,Choi:2002ps,Choi}, which
will be described below.
As the 1-loop correction takes the form:\bea
\frac{1}{8\pi^2}\Delta_a=\sum_{n=0}^{\infty}\int
\frac{d^4l}{(2\pi)^4} \, f_a(p,l,m_n),\eea where $f_a\rightarrow
{1}/{(l^2+m_n^2)^2}$ in the limit $l^2\sim m_n^2\rightarrow \infty$,
 one can introduce a meromorphic pole function:\bea
\label{poleftn}P(q)=\frac{1}{2}\sum_{n=0}^{\infty}\left(\frac{1}{q-m_n}+\frac{1}{q+m_n}\right)\eea
with which  \bea \label{integral}
\frac{1}{8\pi^2}\Delta_a=\int_{\leftrightharpoons} \frac{dq}{2\pi
i}\frac{d^4l}{(2\pi)^4} \, P(q)f_a(p,l,q),\eea where the integration
contour $\leftrightharpoons$ is illustrated as $C_1$ in
Fig.\ref{contour}. This pole function has the following asymptotic
behavior at $|q|\rightarrow \infty$: \bea \label{asymtotic}
P(q)\rightarrow \frac{A}{q}+ B\epsilon({\rm Im}(q))+ {\cal
O}(q^{-2}),\eea where $\epsilon(x)=x/|x|$, and $A$ and $iB$ are real
constants.
With simple dimensional analysis, one easily finds that $A$ and $B$
are associated with logarithmic divergence and linear divergence,
respectively.
In particular, $iB$ corresponds to the spectral density of the KK
spectrum in the  UV limit $m_n\rightarrow \infty$, which is common
to generic 5D field $\Phi(x,y)$ with a definite 4D spin and 4D
chirality, i.e. $\Phi=\phi(x,y)$ or $A_\mu(x,y)$ or
$\psi_{L,R}(x,y)$  with $\gamma_5\psi_{L,R}(x,y)=\pm
\psi_{L,R}(x,y)$.
  As $A$ is
associated with the coefficient (\ref{logd}) of logarithmic
divergence, we have $A\propto (z+z^\prime)$, where $z,z^\prime=\pm
1$ are the orbifold parities of the associated 5D field at
$y=0,\pi$.

One may regulate the 5D momentum integral (\ref{integral}) by
introducing an appropriate set of 5D Pauli-Villars regulator fields
and/or higher derivative regulating terms in the 5D action.
However, as we eventually need to include the gauge boson loops, it
is more convenient to use a dimensional regularization scheme in
which \bea\frac{1}{8\pi^2}\Delta_a&=&\int \frac{d^{D_5}q}{2\pi
i}\frac{d^{D_4}l}{(2\pi)^4} \,
P(q)f_a(p,l,q)  \nonumber \\
&=& \frac{c_a}{8\pi^2}\frac{A}{(4-D_4)} +
\frac{1}{8\pi^2}\Delta_a^\mathrm{finite}, \eea where $c_a$ is some
group theory coefficient, and $\Delta_a^\mathrm{finite}$ is finite
in the limit $D_5\rightarrow 1$ and $D_4\rightarrow 4$. In this
regularization scheme, the irrelevant linear divergence is simply
thrown away, while the logarithmic divergence appears through
$1/(D_4-4)$.

After the integration over $l^\mu$, the remained integration over
$q$ can be done by deforming the integration contour appropriately.
For the 1-loop corrections (\ref{scalar_threshold}) induced by 5D
scalar fields, we find \bea \label{correction}
\frac{1}{8\pi^2}\Delta_a=\int_{C_1} \frac{d^{D_5}q}{2\pi i} P(q)
{\cal G}_a(p,q),\eea where \bea {\cal G}_a(p,q) = \frac{{\rm
Tr}(T_a^2(\phi))}{16\pi^2} \int_0^1 dx (1-2x)^2\left(\frac{2}{4-
D_4} - \ln( q^2 + x(1-x)p^2 )\right).\nonumber\eea Since it depends
on $q^2$ logarithmically, ${\cal G}_a$ contains a branch cut in the
complex plane of $q$, and we can take a branch cut line along the
imaginary axis with $q^2 + x(1-x)p^2 <0$. It is then convenient to
divide the Pole function into three pieces: \bea P(q)= \frac{A}{q} +
B\epsilon(\textrm{Im}q) + P_\mathrm{finite}(q),\eea
 where \bea P_\mathrm{finite}(q)
\rightarrow {\cal O}(q^{-2})\quad {\rm for}\quad |q|\rightarrow
\infty.\nonumber \eea
One can then use the original contour $C_1$ for the integration
involving $B\epsilon({\rm Im} q)$,  an infinitesimal circle around
$q=0$ for the integration involving $A/q$, and finally the contour
deformed as $C_2$ in Fig.\ref{contour} for the integration involving
$P_{\rm finite}$. Applying this procedure to the integral of the
form \bea \Gamma_X=\int_{C_1}\frac{d^{D_5} q}{2\pi i}
\left(\frac{A}{q}+B\epsilon({\rm Im} q)+P_{\rm
finite}\right)\Big(X_0 - \ln (q^2 + X_1^2)\Big), \nonumber\eea one
obtains \bea \left.\Gamma_X\right|_{D_5\rightarrow 1}&=&
A\left(X_0-\ln X_1^2\right)-2iB |X_1|+\int_{C_2}\frac{dq}{2\pi i}
P_{\rm finite}(q)\Big(X_0 - \ln (q^2 + X_1^2)\Big) \nonumber
\\
&=& AX_0 -\ln \left.N(q)\right|_{q=i|X_1|}, \eea where the
$N$-function is defined as \bea \label{N condition} &&P(q) =
\frac{1}{2} \frac{d}{dq}\ln N(q),\nonumber \\&& \frac{1}{2}\ln
N(i|q|) \rightarrow A\ln |q| + iB |q|\quad {\rm for}\quad
|q|\rightarrow \infty.\eea
We then find the 1-loop correction due to a complex 5D scalar field
is given by \bea\label{scalar one loop} \frac{1}{8\pi^2}\Delta_a
=\frac{{\rm Tr}(T_a^2(\phi))}{16\pi^2}\left[\frac{2A}{3(4 -D_4)
}-\int_0^1 dx (1-2x)^2 \ln N(i\sqrt{x(1-x)p^2})\right],
\nonumber\eea showing that the model-parameter dependence of low
energy couplings at $p^2\ll m_{KK}^2$ is determined essentially by
the behavior of $N(q)$ in the limit $q\rightarrow 0$. For a given 5D
gauge or matter field, the corresponding $N(q)$ can be uniquely
determined as will be discussed in Appendix A.

To complete the computation in dimensional regularization, one needs
to subtract the $1/(4-D_4)$ pole  to define the renormalized
coupling.  The subtraction procedure  should take into account that
dimensional regularization has been applied for the momentum
integral defined in the 4D metric frame of $\eta_{\mu\nu}$, while
the correct renormalized coupling should be defined in generic 5D
metric frame as a quantity invariant under the 5D general coordinate
transformation. The $1/(4-D_4)$ pole is associated with the
renormalization of the fixed point gauge couplings, $\kappa_a$ and
$\kappa^\prime_a$, in the action (\ref{action}). For warped
spacetime  with \bea ds^2=e^{2\Omega(y)}\eta_{\mu\nu}dx^\mu
dx^\nu+R^2dy^2\quad (e^{\Omega(0)}=1),\nonumber \eea the logarithmic
divergence structure of (\ref{log div}) indicates that the correct
procedure is to subtract $1/(4-D_4)$ with the counter term
$\lambda_{a0}\ln(\Lambda) + \lambda_{a\pi}\ln(\Lambda
e^{\Omega(y=\pi)})$, which would yield \bea\Delta_a &=&
(\lambda_{a0}+\lambda_{a\pi})\ln\Lambda + \lambda_{a\pi}\ln
\left(e^{\Omega(y=\pi)}\right)+ \Delta_a^\mathrm{finite}. \eea

One can now apply the above prescription  to the 1-loop correction
due to a 5D complex scalar field $\phi_{++}$  on a slice of $AdS_5$:
\bea ds^2=e^{-2k|y|}\eta_{\mu\nu}dx^\mu dx^\nu+R^2dy^2.\nonumber
\eea For $\phi_{++}$ with a 5D mass $M_S\gg p$, there is no zero
mode, and we find \bea \Delta_a (\phi_{++})= \frac{1}{6} {\rm
Tr}(T_a^2(\phi_{++}))\left[\ln\frac{\Lambda}{M_S}-
\frac{1}{2}\ln\left(\frac{\alpha^2-4}{\alpha^2}\right) - \ln
\sinh(\alpha \pi k R) \right], \nonumber \eea where $\alpha =
\sqrt{4 + M_S^2/k^2}$. In fact, one can get the same result using
the Pauli-Villars (PV) regularization scheme in which \bea
\left.\Delta_a(\phi_{++})\right|_{PV} &=& \frac{1}{3} {\rm
Tr}(T_a^2(\phi_{++}))\sum_{n=0}^\infty \ln\left(\frac{m_n(\Phi^{\rm
PV}_{++})}{m_n(\phi_{++})}\right), \eea where $m_n(\Phi^{\rm
PV}_{++})$ is the KK spectrum of the PV regulator field $\Phi^{\rm
PV}_{++}$ which has a bulk mass $\Lambda$. In the limit
$n\rightarrow \infty$, $m_n(\phi_{++})$ and $m_n(\Phi^{\rm
PV}_{++})$ have the same asymptotic form $m_n \rightarrow {n\pi
k}/{(e^{\pi k R} -1)}$. We then have \bea && \hskip
-2cm\left.\Delta_{a}(\phi_{++})\right|_{PV} =\frac{1}{3} {\rm
Tr}(T_a^2(\phi_{++}))\sum_{n=0}^\infty
\left[\,\ln\left(\frac{\Lambda_0}{m_n(\phi_{++})}\right) -
\ln\left(\frac{\Lambda_0}{m_n(\Phi^{\rm
PV}_{++})}\right)\,\right]\nonumber \\
&&\hskip 0.3cm =\left.\Delta_{a}(\phi_{++})\right|_{DR} -
\left.\Delta_{a}(\Phi^{\rm PV}_{++})\right|_{DR} \nonumber \\
&&\hskip 0.3cm = \frac{1}{6} {\rm
Tr}(T_a^2(\phi_{++}))\left[\,\left(\ln\frac{\Lambda}{M_S}-
\frac{1}{2}\ln\left(\frac{\alpha^2-4}{\alpha^2}\right) - \ln
\sinh(\alpha \pi k R)\right)\right. \nonumber \\
&&\hskip 0.7cm+\Lambda \pi R + {\cal O}(1)\Big],\nonumber
 \eea
where $\Lambda_0$ is an arbitrary mass parameter, the subscript DR
means dimensional regularization, and the PV regulator mass is taken
as $\Lambda \gg k, 1/R$. As we have  noticed,  the linearly
divergent part of $\Delta_a$ depends on the employed regularization
scheme,  and such a scheme-dependence can be absorbed into the
renormalization of the Wilsonian 5D gauge couplings. A constant
piece of order unity in $\Delta_a$ is also scheme-dependent, and can
be absorbed into the renormalization of the fixed point gauge
couplings. On the other hand, the terms  depending on the model
parameters $M_S, k, R$ correspond to the calculable part of
$\Delta_a$ which should be scheme-independent.  The above result
confirms that the two regularization schemes, DR and PV, indeed give
the same calculable part of $\Delta_a$.

\section{Warped gauge thresholds}

In this section, we discuss the 1-loop gauge thresholds in generic
5D gauge theory on a slice of AdS$_5$, where some  of the bulk gauge
symmetries are broken by orbifold boundary conditions and/or by bulk
Higgs vacuum values. The effective action of the 4D gauge boson zero
modes $A_{\mu}^{a(0)}$  can be obtained by evaluating \bea
e^{i\Gamma [\Phi_{cl}]} =\int \left[{\cal D}\Phi_{qu}\right]\,
e^{iS_W[\Phi_{cl}+\Phi_{qu}]}, \eea where $\Phi_{cl}$ denotes a
background field configuration which includes $A_{\mu}^{a(0)}$ as
well as the Higgs vacuum values,  and $\Phi_{qu}$ stands for the
quantum fluctuations of all gauge, matter and ghost fields in the
model. To compute the 1-loop effective action, we need the quadratic
action of those quantum fluctuations. To derive the quadratic action
of $\Phi_{qu}$, let us start with the Wilsonian action given by
 \bea \label{5daction} S_W=
S_{\rm gauge}+S_{\rm matter}+S_{\rm gauge-fixing}+S_{\rm ghost},
\eea where \bea &&\hskip -2cm S_{\rm gauge}=-\int d^5x \sqrt{-G}\,
\frac{1}{4}\left(\frac{1}{g_{5A}^2}+\frac{\kappa_A}{4\pi^2}\frac{\delta(y)}{\sqrt{G_{55}}}+\frac{\kappa_A^\prime}{4\pi^2}
\frac{\delta(y-\pi)}{\sqrt{G_{55}}} \right)F^{AMN}F^A_{MN},\nonumber
\\ && \hskip -2cm  S_{\rm matter}= -\int d^5x \sqrt{-G}\,\left[\,\frac{1}{2}
D^M\phi^{I} D_M\phi^I + V(\phi)+i\bar \psi^p(\delta_{pq}\Gamma^M D_M
+{\cal
M}_{Fpq}(\phi)){\psi}^q\right. \nonumber \\
&+&\left.\frac{\delta(y)}{\sqrt{G_{55}}}\left(
V_0(\phi)+2i\mu_{pq}(\phi)\bar\psi^q\psi^q\right)
-\frac{\delta(y-\pi
)}{\sqrt{G_{55}}}\left(V_\pi(\phi)+2i\tilde{\mu}_{
pq}(\phi)\bar\psi^p\psi^q\right)\,\right]\nonumber
  \eea
for the 5D gauge fields $A_M^A$, Dirac fermions $\psi^p$, and real
scalar fields $\phi^I$. Here $S_{\rm gauge-fixing}$ is the
 gauge-fixing term and
 $S_{\rm ghost}$ is the associated ghost action.
We fix the background spacetime to be a slice of AdS$_5$: \bea ds^2
= G_{MN}dx^Mdx^N=e^{-2 k R |y|} \eta_{\mu\nu} dx^\mu dx^\nu + R^2
dy^2 \quad ( 0 \leq y \leq \pi) \nonumber, \eea and impose the
$Z_2\times Z_2^\prime$ orbifold boundary conditions: \bea
\label{oparity}&&
A^{A}_M(-y)=z_A\epsilon_M A^A_M(y),\quad A^A_M(-y+\pi)=z^{\prime}_A\epsilon_M A^A_M(y+\pi),\nonumber \\
&&\phi^I(-y)=z_I\phi^I(y),\quad \phi^I(-y+\pi)=z^{\prime}_I\phi^I(y+\pi),\nonumber \\
&&\psi^p(-y)=z_p\gamma_5\psi^p(y),\quad \psi^p(-y+\pi)=z^{\prime}_
p\gamma_5\psi^p(y+\pi),\eea where  $z_{A,I,p}, z^{\prime}_{A,I,p} =\pm
1, \epsilon_\mu=1$ and $\epsilon_5=-1$. Here we ignore the boundary
kinetic terms of matter fields since they are not relevant for the
discussion of 1-loop gauge couplings. As for the boundary scalar
potentials $V_0$ and $V_\pi$, we assume for simplicity that they
share (approximately) a common minimum with the bulk scalar
potential $V$,  and as a result the scalar field vacuum values are
(approximately) constant along the 5-th dimension: \bea \langle
\phi^I\rangle =v^I. \eea Then there can be two independent sources
of gauge symmetry breaking, one is the bulk Higgs vacuum values
$v^I$ and the other is the orbifold boundary conditions imposed on
the gauge fields.

Let us now set up the notations. In the following, $A_M^\sigma$
denote the 5D gauge fields  {\it not} receiving a mass from the
Higgs vacuum values $v^I$, $B_M^\alpha$ are the other gauge fields
which obtain a nonzero 5D mass,
 $\pi^\alpha$ are the associated  Goldstone
bosons, and finally  $\varphi^i$ are the real-valued physical scalar
field fluctuations in non-Goldstone direction. These gauge and
scalar field fluctuations have the following form of the kinetic and
mass terms:
 \bea \label{convention1}\frac{1}{g_{5A}^2}F^{AMN}F^A_{MN}=\frac{1}{g_{5\sigma}^2}F^{\sigma
MN}F^\sigma_{MN}+\frac{1}{g_{5\alpha}^2}B^{\alpha
MN}B^\alpha_{MN},\nonumber \\
D^M\phi^I D_M\phi^I = \partial^M\varphi^i\partial_M\varphi^i
+\partial^M\pi^\alpha\partial_M\pi^\alpha +\lambda_\alpha^2
B^{M\alpha}B_M^\alpha+...,\nonumber\\ V(\phi)=\langle
V\rangle+\frac{1}{2}{\cal
M}^{2}_{Sij}\varphi^i\varphi^j+...,\nonumber \\
V_0(\phi)=\langle V_{0}\rangle +m_{ij}\varphi^i\varphi^j+...,\quad
V_\pi(\phi)=\langle V_\pi\rangle +\tilde{m}_{ij}\varphi^i\varphi^j+...,
\eea where $F^\sigma_{MN}$ and $B^\alpha_{MN}$ are the field
strength tensor of $A^\sigma_M$ and $B^\alpha_M$, respectively. Here
each 5D field can have arbitrary orbifold parities, and then
$Z_2\times Z^\prime_2$ symmetry implies that the mass matrices take
the form:\bea \label{convention2}
 {\cal M}^{2}_{Sij} = M^{2}_{Sij}
\epsilon_{z_{ij} z_{ij}'}(y),\quad
 {\cal M}_{Fpq} = M_{Fpq}
\epsilon_{\bar z_{pq} \bar z_{pq}'}(y),\nonumber \\
m_{ij} = m_{ij}\delta_{z_iz_j},\quad
\tilde{m}_{ij} = \tilde{m}_{ij}\delta_{z'_iz'_j},\nonumber \\
\mu_{pq} = \mu_{pq}\delta_{\bar z_pz_q},\quad \tilde{\mu}_{pq} =
\tilde{\mu}_{pq}\delta_{\bar z'_pz'_q},\eea
 where $M^{2}_{Sij}$ and
$M_{Fpq}$ are constant,
 $z_{ij} =
z_i z_j $, $z'_{ij} = z'_i z'_j $, e.t.c., $\bar{z}=-z$,
$\bar{z}^\prime=-z^\prime$, and the kink function
$\epsilon_{zz'}(y)$ is defined as \bea \hskip -1cm \epsilon_{zz'}(y)
= 1\,\,\, {\rm for} \,\,\, 0 < y < \pi, \quad  \epsilon_{z z'}(-y) =
z \epsilon_{z z'}(y), \quad \epsilon_{z z'}(\pi  - y) = z'
\epsilon_{z z'}(\pi +y). \nonumber \eea Note that $M_{Sij},
M_{Fpq},m_{ij}$ and $\tilde{m}_{ij}$ have the mass dimension one,
while $\mu_{pq}$ and $\tilde{\mu}_{pq}$ are dimensionless
parameters. For a generic form of mass matrices,  there can be
nonzero mass mixing between matter fields with different orbifold
parities. Our aim is to compute the 1-loop gauge couplings  as a
function of the mass parameters and the orbifold parities, which are
defined above.

As we are going to compute the low energy effective action of
$A_\mu^{a(0)}(x)$, we regard all 5D gauge fields as quantum
fluctuations around a background configuration  of
 $A_\mu^{a(0)}(x)$ which correspond to the zero modes of $\left.A_M^{\sigma}\right|_{\sigma=a}$  having  the orbifold parity $z=z^\prime=1$. To
proceed,  we choose the following form of the gauge fixing term:\bea
\hskip -2cm S_\mathrm{gauge-fixing} &=& -\int d^5x
\sqrt{-G}\left[\frac{1}{2g_{5\sigma}^2} \left(e^{2kR|y|}
\eta^{\mu\nu}D_\mu A^{\sigma}_{\nu}
+ \frac{1}{R^2}e^{2k R|y|}\partial_y(e^{-2k R|y|}A^\sigma_{5})\right)^2\right.  \nonumber \\
 &+& \left.\frac{1}{2g_{5\alpha}^2} \left(e^{2kR|y|}
\eta^{\mu\nu}D_\mu B^{\alpha}_{\nu} + \frac{1}{R^2}
e^{2kR|y|}\partial_y(e^{-2kR|y|}B^\alpha_{5}) - g_{5\alpha}^2
\lambda_\alpha  \pi^\alpha\right)^2\right], \eea where
 $D_\mu =
\partial_\mu - i A^{a(0)}_{\mu}T^a$ is the covariant
derivative involving the background gauge boson zero modes.
 The corresponding ghost action is given by \bea \hskip -0.5cm S_{\rm
ghost}&=&\int d^5x \sqrt{-G}\left[\,e^{2kR|y|} \bar
c^\sigma_AD^2c^\sigma_A + \frac{e^{2kR|y|}}{R^2} \bar c^\sigma_A
(\partial_y e^{-2kR|y|}\partial_y c^\sigma_A) \right.\nonumber \\
 &+& \left. e^{2kR|y|} \bar c^\alpha_B D^2c^\alpha_B +
\frac{e^{2kR|y|}}{R^2} \bar c^\alpha_B (\partial_y
e^{-2kR|y|}\partial_y c^\alpha_B) -  g^2_{5\alpha}\lambda^2_\alpha
\bar c_B^\alpha c_B^\alpha+...\,\right], \eea where $c_A^\sigma$ and
$c_B^\alpha$ are the ghost fields for $A^\sigma_M$ and  $B^i_M$,
respectively, and $D^2=\eta^{\mu\nu}D_\mu D_\nu$.

In the model under consideration, there are three class of field
fluctuations, each of which can have arbitrary orbifold parities:
(i) 5D gauge fields $A_M^\sigma$ which do not get a mass from the
Higgs vacuum values $v^I$, and the associated ghost fields
$c_A^\sigma$, (ii) 5D gauge fields $B_M^\alpha$ which get a nonzero
5D mass $M_{V\alpha}=g_{5\alpha}\lambda_\alpha$ from $v^I$, and the
associated Goldstone bosons and ghost fields, $\pi^\alpha$ and
$c_B^\alpha$, (iii) 5D Dirac fermions  $\psi^p$ and the physical
scalar fields $\varphi^i$.
 After the following  field
redefinition, \bea\label{redef} A^\sigma_M \rightarrow \frac{1}{\sqrt{R}}
g_{5\sigma} A^\sigma_M ,\quad B^\alpha_M \rightarrow
\frac{1}{\sqrt{R}} g_{5\alpha} B^\alpha_M,\quad \pi^\alpha\rightarrow \frac{1}{\sqrt{R}}\pi^\alpha\nonumber \\
\varphi^i\rightarrow \frac{1}{\sqrt{R}} \varphi^i,\quad \psi^p
\rightarrow \frac{1}{\sqrt{R}}\psi^p, \quad c^\sigma_A \rightarrow
\frac{1}{\sqrt{R}} c^\sigma_A,\quad c^\alpha_B \rightarrow
\frac{1}{\sqrt{R}} c^\alpha_B, \eea we find that each class of field
fluctuations has the quadratic action: \bea \label{quad} S_2=\int
d^4x dy \left(\,{\cal L}_A + {\cal L}_B +{\cal L}_M\,\right),\eea
where \bea {\cal L}_A &=& -\frac{1}{2}\eta^{\mu\nu} A_\mu^\sigma
\Delta A^\sigma_\nu - \frac{e^{-2kR|y|}}{2R^2} (\partial_y
A^\sigma_\mu)^2 - \frac{e^{-2kR|y|}}{2} A_5^\sigma \Delta A^\sigma_5
- \frac{e^{-4kR|y|}}{2R^2} (\partial_y  A^\sigma_5)^2 \nonumber \\
\hskip 0cm &-&  \frac{1}{2}  \frac{ e^{-4kR|y|}}{R}\left(-4k^2 +
4k (\delta(y)-\delta(y-\pi )\right)(A^\sigma_5)^2 \nonumber \\
\hskip 0cm &-&   e^{-2kR|y|} \bar c^\sigma_A\Delta c^\sigma_A
+ \frac{e^{-2kR|y|}}{R^2} \bar c^\sigma_A (\partial_y  e^{-2kR|y|}\partial_y c^\sigma_A)  \nonumber \\
\hskip 0cm {\cal L}_B &=& -\frac{1}{2}  \eta^{\mu\nu}B^\alpha_\mu
\Delta B^\alpha_{\nu} - \frac{e^{-2kR|y|}}{2R^2}(\partial_y
B^\alpha_\mu)^2
 -\frac{e^{-2kR|y|}}{2}g^2_{5\alpha}  \lambda^2_\alpha (B^\alpha_\mu)^2 \nonumber \\
\hskip 0cm  &-& \frac{1}{2}  e^{-2kR|y|}\left( \pi^\alpha \Delta
\pi^\alpha + B_5^\alpha \Delta B_5^\alpha \right) -
\frac{e^{-4kR|y|}}{2R^2}\left((\partial_y \pi^\alpha)^2
+ (\partial_y B_5^\alpha)^2 \right) \nonumber \\
\hskip 0cm &-&  \frac{1}{2}e^{-4kR|y|} \left(\begin{array}{cc}
\pi^\alpha & B^\alpha_{5}
\end{array}\right) \left(\begin{array}{cc} g^2_{5\alpha}\lambda^2_\alpha & -2k
g_{5\alpha}\lambda_\alpha \epsilon_{--}(y) \\ -2k
g_{5\alpha}\lambda_\alpha \epsilon_{--}(y) &
g^2_{5\alpha}\lambda^2_\alpha - 4k^2
\end{array}\right)
\left(\begin{array}{c}  \pi^\alpha \\ B^\alpha_{5} \end{array}\right) \nonumber\\
\hskip 0cm &-&
\mathinner{ \frac{e^{-4kR|y|}}{R}}  2k (\delta(y)-\delta(y-\pi ))( B^\alpha_5)^2 \nonumber \\
\hskip -2.5cm &-&   e^{-2kR|y|} \bar c^\alpha_B\Delta c^\alpha_B +
\frac{e^{-2kR|y|}}{R^2} \bar c^\alpha_B (\partial_y
e^{-2kR|y|}\partial_y c^\alpha_B)
 -e^{-4kR|y|} g^2_{5\alpha}\lambda^2_\alpha \bar c_B^\alpha c_B^\alpha \nonumber \\
\hskip 0cm {\cal L}_M &=& -  \frac{1}{2} e^{-2kR|y|}\varphi^i \Delta
\varphi^i - \frac{e^{-4kR|y|}}{2R^2}(\partial_y \varphi^i)^2
- \frac{e^{-4kR|y|}}{2}{\cal M}^2_{S ij}\varphi^i\varphi^j \nonumber \\
\hskip 0cm  &-&  \frac{e^{-4kR|y|}}{R}\left(
m_{ij}\delta(y)
- \tilde m_{ij}\delta(y-\pi )\right)\varphi^i\varphi^j \nonumber \\
\hskip 0cm &-&  i  e^{-3kR|y|}\bar\psi^p \gamma^\mu D_\mu \psi^p -
i\frac{e^{-2kR|y|}}{R}\bar\psi^p\gamma_5(\partial_y
e^{-2kR|y|}\psi^p)
- i e^{-4kR|y|}{\cal M}_{Fpq}\bar\psi^p\psi^q \nonumber \\
\hskip 0cm &-&  i \frac{e^{-4kR|y|}}{R}\left(2\mu_{pq}\delta(y) -
2\tilde\mu_{pq}\delta(y-\pi )\right)\bar\psi^p\psi^q. \nonumber\eea
Here the gauge-covariant operator $\Delta$ is defined as
\bea\Delta\Phi=\left(- \eta^{\mu\nu}D_\mu D_\nu +
F^{(0)}_{\mu\nu}{\cal J}_j^{\mu\nu}\right)\Phi,\eea
 where
$F^{(0)}_{\mu\nu}=F^{a(0)}_{\mu\nu}T^a$ is the field strength of the
gauge boson zero modes $A_\mu^{a(0)}$, and ${\cal J}_j^{\mu\nu}$ is
the 4D Lorentz generator for a field with 4D spin $j$, which is
normalized as ${\rm tr}({\cal J}^{\mu\nu}_j{\cal
J}^{\rho\lambda}_j)=C(j)(\eta^{\mu\rho}\eta^{\nu\lambda}-\eta^{\mu\lambda}\eta^{\nu\rho})$,
where $C(j)=(0,1/2,2)$ for $j=(0,1/2,1)$. Here and in the following,
 $\Phi(x,y)$ stands for a 5D field which has
a definite value of 4D spin $j$ and also of 4D chirality,  e.g.
$\Phi=A_\mu(x,y)$ with $j=1$,  $\Phi=\psi_{L,R}(x,y)$ with $j=1/2$
and $\gamma_5\psi_{L,R}=\pm\psi_{L,R}$, $\Phi=A_5(x,y)$ or
$\varphi(x,y)$ or $c_A(x,y)$ with $j=0$.
   Note that the AdS curvature
$k$ generates a mixing between $B^\alpha_5$ and $\pi^\alpha$ in the
quadratic action. Since the Goldstone boson $\pi^\alpha$ has the
same orbifold parity as $B^\alpha_\mu$, this is a mixing between 4D
scalar fields  with opposite orbifold parities.

With the quadratic action (\ref{quad}), the 1-loop effective action
of the gauge boson zero modes is given by \bea \Gamma[A^{(0)}_{\mu}]
&=&
 - \frac{\pi R}{4g_{5a}^ 2}\int d^4x\,
F^{a(0)}_{\mu\nu}F^{a(0)\mu\nu}\nonumber \\
&+&\frac{i}{2} \left(\textrm{Tr}_{A_\mu}\ln \tilde\Delta_{A_\mu} +
\textrm{Tr}_{A_5}\ln \tilde\Delta_{A_5}
- 2\textrm{Tr}_{c_A}\ln\tilde\Delta_{c_A}  \right)  \\
&+& \frac{i}{2} \left(\textrm{Tr}_{B_\mu}\ln \tilde\Delta_{B_\mu} +
\textrm{Tr}_{B_5,\pi}\ln \tilde\Delta_{B_5,\pi}
- 2\textrm{Tr}_{c_B}\ln\tilde\Delta_{c_B}  \right) \nonumber \\
&+& \frac{i}{2} \left(\textrm{Tr}_{\varphi}\ln \tilde
\Delta_{\varphi} - \textrm{Tr}_{\psi_L}\ln
\tilde\Delta_{\psi_L}-\textrm{Tr}_{\psi_R}\ln
\tilde\Delta_{\psi_R}\right), \nonumber \eea where \bea
\textrm{Tr}_\Phi\ln \tilde \Delta_\Phi &=& \sum_n
\textrm{Tr}_{\Phi_n}\ln\left(\Delta +
m^2_n(\Phi)\right)\nonumber \\
&=&\int_{\leftrightharpoons}\frac{ d^{D_5}q}{2\pi i} P_\Phi(q)
\textrm{Tr}\ln\left(\Delta + q^2\right).
  \eea Here $\Phi_n$
denotes the $n$-th KK modes  with the mass eigenvalue $m_n(\Phi)$:
\bea \Phi(x,y)=\sum_n \Phi_n(x) f_n(y),\eea and in the last step we
have applied the Pole function technique discussed in the previous
section:
 \bea P_\Phi(q) =\frac{1}{2}\sum_n\left(\frac{1}{q-m_n(\Phi)}+\frac{1}{q+m_n(\Phi)}\right),\eea
where the summation includes the zero modes also.

It is straightforward to perform the integration over 4D loop
momentum with dimensional regularization. We then find \bea
\textrm{Tr}\ln\left(\Delta + q^2\right) = i \int \frac{d^4
p}{(2\pi)^4} {\cal G}^\Phi_a(q,p)A^a_\mu(-p)(p^2 \eta^{\mu\nu} -
p^\mu p^\nu)A^a_\nu(p)+ \cdots, \nonumber \eea where \bea {\cal
G}^\Phi_a(q,p) = \frac{1}{8\pi^2}{\rm Tr}(T_a(\Phi)^2) \int_0^1 dx
\left(\frac{1}{2} d(j_\Phi)(1-2x)^2
- 2 C(j_\Phi)\right)\nonumber \\
\quad\quad \times \left(\frac{2}{4-D_4} + \ln(4\pi e^{-\gamma}) -
\ln(q^2 + x(1-x)p^2)\right). \eea  Here $d(j_\Phi)=(1, 2, 2, 4)$ and
$C(j_\Phi)=(0, 1/2, 1/2, 2)$ for $j_\Phi=(0,1/2_L, 1/2_R, 1)$ denoting the 4D spin and
chirality
of $\Phi$.
The 1-loop correction induced by $\Phi$ can be expressed as \bea
\frac{1}{8\pi^2}\Delta_a^\Phi(p) = (-1)^{2j_\Phi}\int_{\leftrightharpoons} \frac{
d^{D_5}q}{2\pi i} P_\Phi(q) {\cal G}^\Phi_a(q,p), \eea where the
dependence on the 4D spin and unbroken gauge charges of $\Phi$ is
encoded in ${\cal G}_a^\Phi$, while the dependence on various mass
parameters is encoded in the pole function $P_\Phi$ which contains
the full information on the KK spectrum. As explained in Section
2, we can deform the integration contour appropriately to simplify
the integration over $q$. (See Fig. 1.)
Then, following the method discussed in the previous section, we
find \bea \label{master_eq} \frac{1}{8\pi^2}\Delta_a^\Phi(p) &=&
\frac{(-1)^{2j_\Phi}}{8\pi^2}{\rm Tr}(T_a^2(\Phi)) \left\lbrace \left(\frac{1}{12}
d(j_\Phi)- C(j_\Phi)  \right)
\left(\frac{2A_\Phi}{4-D_4}+{\cal O}(1)\right)  \right. \nonumber \\
&&\hskip -2cm+ \left. \int_0^1 dx \left( C(j_\Phi)-\frac{1}{4}
d(j_\Phi) (1-2x)^2 \right) \ln N^\Phi (i \sqrt{x(1-x)p^2})
\right\rbrace, \eea where the Pole function has the following
asymptotic behavior  at $|q|\rightarrow \infty$: \bea P_\Phi(q)
= \frac{A_\Phi}{q} + B_\Phi \epsilon(\textrm{Im}q)+ {\cal
O}\left(q^{-2}\right)\eea and
 $N^\Phi$ is a holomorphic even function define as
\bea N^{\Phi}=C_\Phi\prod_n \left(m_n^2(\Phi)-q^2\right) \qquad
\left(\,P_\Phi(q)= \frac{1}{2} \frac{d}{dq}\ln N^\Phi(q)\,\right),
\nonumber
\\
\frac{1}{2}\ln N^\Phi(i|q|)= A_\Phi\ln|q| +iB_\Phi |q| +{\cal
O}(|q|^{-1}) \quad {\rm at} \quad |q|\rightarrow \infty.\eea

Since $A_\Phi/(4-D_4)$ is associated with the logarithmic divergence
of the fixed point gauge couplings, we have $A_\Phi\propto
(z+z^\prime)$, where $z,z^\prime$ are the orbifold parities of
$\Phi$. (See Eqs.(\ref{log div}) and (\ref{logd}).) In our
convention, for $\Phi(x,y)=\{\phi, \psi_L, \psi_R, A_\mu\}$,
we have
\bea A_\Phi=\frac{z+z^\prime}{4},\quad iB_\Phi=\frac{e^{\pi kR}-1}{2
k},\eea
 where $\Phi(-y)=z\Phi(y)$ and
$\Phi(-y+\pi)=z^\prime\Phi(y+\pi)$. Note that here $\phi$ can be a
5D scalar, or the 5-th component of a 5D vector, or a ghost field.
Also a 5D Dirac fermion $\psi$ with orbifold parities $z,z^\prime$
consists of $\psi_L$ with orbifold parities $z,z^\prime$ and
$\psi_R$ with orbifold parities
$\bar{z}=-z,\bar{z}^\prime=-z^\prime$, and thus
$A_\psi=A_{\psi_L}+A_{\psi_R}=0$.
 As was noticed in the previous
section, in warped spacetime, the renormalized fixed point gauge
couplings at the cutoff scale $\Lambda$ are obtained by subtracting
the pole divergence $(z+z^\prime)/(4-D_4)$ with a counter term
proportional to \bea {\delta(y)}z\ln\Lambda +{\delta(y-\pi)}
z^\prime\ln (e^{-k\pi R}\Lambda).\nonumber \eea

We are now ready to present the 1-loop corrections to low energy
gauge couplings, induced by  generic 5D fields on a slice of
AdS$_5$. For this, let $N^\Phi_{zz^\prime}$ denote the $N$-function
of $\Phi(x,y)$ having a definite value of 4D spin $j_\Phi$, of 4D
chirality, and of orbifold parities $z,z^\prime$.
Explicit forms of $N^{\Phi}_{zz^\prime}(q)$ and their limiting
behaviors at $|q|\rightarrow 0, \infty$  for $\Phi$'s with generic
bulk and boundary masses are presented in Appendix A.
 Also let $\{\Phi\}$ denote a set of $\Phi$'s having the same $j_\Phi$ and unbroken gauge charges, but not necessarily
the same orbifold parities, which generically have a mixing to each
other in the quadratic action (\ref{quad}) of quantum fluctuations,
and $N^{\{\Phi\}}$ denote the $N$-function of this set of $\Phi$'s.
Then the full 1-loop corrections are summarized as \bea
\frac{1}{8\pi^2} \Delta_a=
\frac{1}{8\pi^2}\left[\,\Delta^{\{A\}}_a+\Delta^{\{B\}}_a+\Delta^{\{\psi_L\}}_a+\Delta^{\{\psi_R\}}_a+
\Delta^{\{\varphi\}}_a\,\right], \eea where \bea\label{form}
\Delta^{\{A\}}_a&=&\Delta_a^{\{A_\mu\}}(p) + \Delta_a^{\{A_5\}}(p)
- 2\Delta_a^{\{c_A\}}(p), \nonumber \\
\Delta^{\{B\}}_a&=&
  \Delta_a^{\{B_\mu\}}(p)+\Delta_a^{\{B_5,\pi\}}(p)
- 2\Delta_a^{\{c_B\}}(p)\eea for $\Delta_a^{\{\Phi\}}(p)$ given by
\bea \label{boson_th}\Delta^{\{\Phi\}}_a(p) = {a(j_\Phi)} {\rm
Tr}(T_a^2(\Phi)) \left[ n^{\{\Phi\}}_{0} \ln p
-\frac{1}{2}\left(n^{\{\Phi\}}_{++}-n^{\{\Phi\}}_{--}\right)\ln\Lambda
\right. \nonumber \\ \hskip 1.5cm +\left.\frac{1}{2}\ln
N_{0}^{\{\Phi\}} + \frac{1}{4}(n^{\{\Phi\}}_{++} - n^{\{\Phi\}}_{+-}
+ n^{\{\Phi\}}_{-+}-n^{\{\Phi\}}_{--})\pi k R+ {\cal
O}\left(1\right) \right].\eea Here $n^{\{\Phi\}}_0$ denotes the
number of zero modes in $\{\Phi\}$, $n^{\{\Phi\}}_{zz^\prime}$ is
the number of $\Phi$'s with orbifold parities $z,z^\prime$ defined
as \bea \Phi(-y)=z\Phi(y),\quad
\Phi(-y+\pi)=z^\prime\Phi(y+\pi),\nonumber \eea and
  \bea a(j_\Phi)=
\left( -\frac{1}{6},-\frac{2}{3}, \frac{10}{3} \right) \quad {\rm
for} \quad j_\Phi=\left( 0, \frac{1}{2},
 1\right),\nonumber \\
 N^{\{\Phi\}}(q) =
(-q^2)^{n_0^{\{\Phi\}}}\left(N_{0}^{\{\Phi\}} + {\cal
O}(q^2/m_{KK}^2)\right),\eea where $m_{KK}$ denotes the lightest KK
mass of $\{\Phi\}$.
The above result shows that the model-parameter dependence of 1-loop
gauge couplings is determined  mostly by the behavior of
$N$-functions at $|q|\ll m_{KK}$, particularly  by $N_0^{\{\Phi\}}$.

The 1-loop corrections induced by 5D Dirac fermions $\{\psi^p\}$
take a simpler form.   As the equation of motion for $\psi$ involves
$\gamma_5$, it is convenient to split each $\psi$ into two chiral
fermions: $\psi=\psi_L+\psi_R$ with $\gamma_5\psi_{L,R}=\pm
\psi_{L,R}$, and then we always have \bea
n_{zz^\prime}^{\{\psi_L\}}=n_{\bar{z}\bar{z}^\prime}^{\{\psi_R\}},\quad
n_0^{\{\psi_L\}}-n_0^{\{\psi_R\}}=n^{\{\psi_L\}}_{++}-n^{\{\psi_R\}}_{++},\eea
regardless of
 the bulk and boundary fermion masses  ${\cal
M}_{Fpq},\mu_{pq}$ and $\tilde{\mu}_{pq}$. (If there is no mass
mixing between $\psi^p$ with different orbifold parities,
$n_0^{\{\psi_L\}}=n^{\{\psi_L\}}_{++}$ and
$n_0^{\{\psi_R\}}=n^{\{\psi_R\}}_{++}=n^{\{\psi_L\}}_{--}$.) Since
$\{\psi_L\}$ and $\{\psi_R\}$ have the same KK mass spectrum, we
also have  \bea
N^{\{\psi_L\}}(q)=(-q^2)^{n_0^{\{\psi_L\}}-n_0^{\{\psi_R\}}}
N^{\{\psi_R\}}(q)\eea and thus \bea
N^{\{\psi\}}(q)&=&N^{\{\psi_L\}}(q)N^{\{\psi_R\}}(q)\nonumber \\
&=&
(-q^2)^{n_0^{\{\psi_L\}}+n_0^{\{\psi_R\}}}\left[\left(N^{\{\psi_L\}}_0\right)^2+{\cal
O}(q^2/m_{KK}^2)\right], \eea where \bea
N^{\{\psi_L\}}(q)&=&(-q^2)^{n_{0}^{\{\psi_L\}}}\left({N^{\{\psi_L\}}_0}+{\cal
O}(q^2/m_{KK}^2)\right) \nonumber \eea in the limit $|q|\ll m_{KK}$.
We then find  the 1-loop gauge couplings induced by $\{\psi\}$ are
given by
\bea
\label{fermion_th}\frac{1}{8\pi^2}\Delta^{\{\psi\}}_a(p)=\frac{1}{8\pi^2}\left(\Delta^{\{\psi_L\}}_a(p)+\Delta^{\{\psi_R\}}_a(p)\right)
\nonumber \\= -\frac{1}{12\pi^2}{\rm Tr}(T^2_a(\psi)) \left[
\left(n_{0}^{\{\psi_L\}} + n_{0}^{\{\psi_R\}} \right)\ln p + \ln
N_{0}^{\{\psi_L\}} + {\cal O}(1) \right].\eea

In the above, the external momentum $p^\mu$ of the gauge boson zero
mode is assumed to be smaller than the lowest KK mass, justifying
the use of the $N$-function at $q\rightarrow 0$. However, in certain
parameter limit, there might be a KK state having a particularly
light mass. For instance, the lightest KK state of a Dirac fermion
$\psi_{+-}$ with bulk mass $M_{F}> k$ has a 4D mass
$m^{\psi}_{KK}\sim k e^{-(k+2M_F)\pi R/2}$ which can be much smaller
than  1 TeV even when $ke^{-k\pi R}\gtrsim {\cal O}(1)$ TeV. In such
case, one needs to consider the gauge couplings at $p>
m^{\psi}_{KK}$, which can be easily obtained from (\ref{boson_th}).
To see this, let us consider the case with \bea m_1(\Phi) \leq
m_2(\Phi)\leq ...\leq m_n(\Phi) < p < m_{n+1}(\Phi), \eea in which
there are $n_0^\Phi+n$ light modes with a mass smaller than $p$.
 One can then consider the
$N$-function at $m_n < q < m_{n+1}$, which can be expressed as \bea
N^\Phi(q)=(-q^2)^{n_0^\Phi}\left(\prod_{l=1}^{n}(m_l^2-q^2)\right)\left(N_{n}^\Phi+{\cal
O}(q^2/m_{n+1}^2)\right),\eea where \bea
N_{n}^\Phi=N_0^\Phi/\prod_{l=1}^{n} m_l^2,\eea and find that the
1-loop gauge couplings at $m_n< p<m_{n+1}$ are given by \bea
\frac{1}{8\pi^2}\Delta^{\Phi}_a(p) = \frac{a(j_\Phi)}{8\pi^2} {\rm
Tr}(T_a^2(\Phi)) \left[ \left(n^{\Phi}_{0}+n\right) \ln p
-\frac{1}{2}\left(n^{\Phi}_{++}-n^{\Phi}_{--}\right)\ln\Lambda
\right. \nonumber
\\ \hskip 1.5cm +\left.\frac{1}{2}\ln N_n^{\Phi} +
\frac{1}{4}(n^{\Phi}_{++} - n^{\Phi}_{+-} +
n^{\Phi}_{-+}-n^{\Phi}_{--})\pi k R+ {\cal O}\left(1\right)
\right].\eea

 As the $N$-functions play a crucial role
in our analysis, let us discuss some relevant features of
$N^{\{\Phi\}}$ here. More complete discussion will be given in
Appendix A. First, for $A^\sigma_{M}=(A^\sigma_\mu,A^\sigma_5)$ and
$c^\sigma_A$ with the orbifold parities \bea
A_\mu^\sigma(-y)=z_\sigma A_\mu^\sigma(y),\quad
A_5^\sigma(-y)=-z_\sigma A_5^\sigma(y)\equiv \bar{z}_\sigma
A_5^\sigma(y),\nonumber \\
A_\mu^\sigma(-y+\pi)=z^\prime_\sigma A_\mu^\sigma(y+\pi),\quad
A_5^\sigma(-y+\pi)=-z^\prime_\sigma A_5^\sigma(y+\pi)\equiv
\bar{z}^\prime_\sigma
A_5^\sigma(y+\pi),\nonumber \\
c_A^\sigma(-y)=z_\sigma c_A^\sigma(y),\quad
c_A^\sigma(-y+\pi)=z^\prime_\sigma c_A(y+\pi)\nonumber \eea
 we have \bea N^{A^\sigma_\mu}_{z_\sigma z_\sigma^\prime}(q) = N^{c^\sigma_A}_{z_\sigma z_\sigma^\prime}(q)
= (-q^2)^{(z_\sigma+z_\sigma')/2}
N^{A^\sigma_5}_{\bar{z}_\sigma\bar{z}_\sigma^\prime}(q). \eea
This relation simply means that $A_\mu^{\sigma}$, $A_5^{\sigma}$ and
$c_A^{\sigma}$ have the same KK mass spectra, which explains the
form of $\Delta_a^{\{A\}}$ in (\ref{form}). Here the factor
$q^{z_\sigma+z_\sigma^\prime}$ represents the zero mode of
$A_{\mu}^\sigma$ with $z_\sigma=z_\sigma^\prime=1$ or of
$A_{5}^{\sigma}$ with $\bar{z}_\sigma=\bar{z}_\sigma^\prime=1$. In
the quadratic action (\ref{quad}), $A_M^\sigma$ does not have any
mixing with other fields, and therefore \bea
N^{\{A_\mu\}}=\prod_\sigma N_{z_\sigma
z_\sigma^\prime}^{A_\mu^\sigma},\quad N^{\{A_5\}}=\prod_\sigma
N_{\bar z_\sigma \bar z_\sigma^\prime}^{A_5^\sigma},\quad
N^{\{c_A\}}=\prod_\sigma N_{z_\sigma z_\sigma^\prime}^{c_A^\sigma}.
 \eea

  On the other hand, for $B^\alpha_M=(B^\alpha_\mu, B^\alpha_5)$ and the associated Goldstone and ghost fields,
  $\pi^\alpha$ and
$c^\alpha_B$, there is a mass mixing between $B^\alpha_5$ and
$\pi^\alpha$ which have opposite orbifold parities. We still have
\bea N^{B^\alpha_\mu}_{z_\alpha z_\alpha^\prime}(q) =
N^{c^\alpha_B}_{z_\alpha z_\alpha^\prime}(q), \quad N^{\lbrace
B^\alpha_5, \pi^\alpha \rbrace}(q)= N^{B^\alpha_\mu}_{z_\alpha
z_\alpha^\prime}(q) N^{\tilde{B}^\alpha_\mu}_{\bar z_\alpha\bar
z_\alpha^\prime}(q) , \eea where $\tilde{B}^\alpha_\mu$ is an
artificial vector field which has the same bulk mass as
$B^\alpha_\mu$ and also the boundary masses given by\bea -\int d^4x
dy \frac{e^{-2 k R
|y|}}{R}\Big(2k\delta(y)-2k\delta(y-\pi)\Big)\eta^{\mu\nu}\tilde
B_\mu^\alpha \tilde B^{\alpha}_\nu.\nonumber\eea We then find \bea
&& N^{\{B_\mu\}}=\prod_\alpha N^{B_\mu^\alpha}_{z_\alpha
z^\prime_\alpha},\quad N^{\{c_B\}}=\prod_\alpha
N^{c_B^\alpha}_{z_\alpha z^\prime_\alpha},\nonumber \\
&& N^{\lbrace B_5, \pi\rbrace}(q)= \prod_\alpha\left(
N^{B^\alpha_\mu}_{z_\alpha z_\alpha^\prime}(q)
N^{\tilde{B}^\alpha_\mu}_{\bar z_\alpha\bar
z_\alpha^\prime}(q)\right). \eea


In the presence of mixing between fields with different orbifold
parities,  $N^{\{\psi\}}$ and $N^{\{\varphi\}}$ generically take a
highly complicate form. Here we present the results  for relatively
simple cases, (i) two Dirac fermions with generic bulk and boundary
masses and (ii) two scalar fields with just bulk masses,  while
leaving the discussion for more general case in Appendix A. Let us
first consider the case of two Dirac fermions
 $\lbrace \psi^p_{z_p z'_p} \rbrace$  ($p=1,2$) with the following bulk and boundary masses:
\bea {\cal M}_{F pq},\quad  \mu_{12},\quad \tilde\mu_{12}.\eea Note
that $\mu_{pp}=\tilde\mu_{pp}=0$, and the Dirac fermion $\psi^p_{z_p
z^\prime_p}$ consists of $\psi_L^p$ with orbifold parities
$z_p,z_p^\prime$ and $\psi_R^p$ with orbifold parities $\bar{z}_p=
-z_p, \bar{z}_p^\prime= -z_p^\prime$.
 In
the fundamental domain $0<y<\pi$, the $2\times 2$ bulk mass matrix
can be described by two
 mass eigenvalues $M_{Fp}$ ($p=1,2$) and a mixing angle $\theta_F$:
  \bea \hskip -1cm U \left(\begin{array}{cc} M_{F11} & M_{F12}
\\ M_{F21} &  M_{F22}\end{array} \right)U^\dagger =
\left(\begin{array}{cc} M_{F1} & 0 \\ 0 & M_{F2} \end{array}
\right), \quad U= \left(\begin{array}{cc} \cos\theta_F &
\sin\theta_F
\\-\sin\theta_F & \cos\theta_F\end{array} \right). \eea Let $N_{zz^\prime}^{\psi_{L,R}(M)}$ denote the $N$ function of
$\psi_{L,R}$ with orbifold parities $z,z^\prime$ and a bulk mass
$M$. We then find that the $N$-function of the above two Dirac
fermions is given by \bea N^{\{\psi^1,\psi^2\}}(q) &=& N^{\{\psi_L^1,\psi_L^2\}}(q) N^{\{\psi_R^1,\psi_R^2\}}(q)\nonumber \\
&=& (-q^2)^{-(z_1+z_1'+z_2+z_2')/2}
\left(N^{\{\psi_L^1,\psi_L^2\}}(q)\right)^2, \eea where \bea
\label{two fermions} \hskip -2cm N^{\lbrace \psi_L^1,\psi_L^2
\rbrace}(q) = \left(c_0 c^*_\pi N^{\psi_L(M_{F1})}_{z_1z'_1} + s_0
s^*_\pi N^{\psi_L(M_{F2})}_{z_1z'_1}\right)
\left(s^*_0 s_\pi N^{\psi_L(M_{F1})}_{z_2z'_2} + c^*_0 c_\pi N^{\psi_L(M_{F2})}_{z_2z'_2}\right) \nonumber\\
- \left(c_0 s_\pi N^{\psi_L(M_{F1})}_{z_1z'_2} - s_0 c_\pi
N^{\psi_L(M_{F2})}_{z_1z'_2}\right) \left(s^*_0 c^*_\pi
N^{\psi_L(M_{F1})}_{z_2z'_1} - c^*_0 s^*_\pi
N^{\psi_L(M_{F2})}_{z_2z'_1}\right) \eea for \bea \label{cs}
c_{0}
&=& \frac{\cos\theta_F -
z_1\mu_{12}\sin\theta_F}{\sqrt{1+|\mu_{12}|^2}},
\quad c_{\pi} = \frac{\cos\theta_F - z'_1\tilde{\mu}_{12}\sin\theta_F}{\sqrt{1+|\tilde\mu_{12}|^2}}\nonumber \\
s_{0} &=& \frac{\sin\theta_F
+z_1\mu_{12}\cos\theta_F}{\sqrt{1+|\mu_{12}|^2}}, \quad s_{\pi} =
\frac{\sin\theta_F +
z'_1\tilde{\mu}_{12}\cos\theta_F}{\sqrt{1+|\tilde\mu_{12}|^2}}.
\eea Note that
 this $N$-function takes a
factorized form,
$N^{\{\psi_L^1,\psi_L^2\}}=N^{\psi_L(M_{F1})}_{zz'}N^{\psi_L(M_{F2})}_{zz'}$
if $\psi^1$ and $\psi^2$ have the same orbifold parities. One can
similarly get the $N$-function of two scalar field system $\lbrace
\varphi^i_{z_i z'_i}\rbrace$ ($i=1,2$) with a generic form of the
bulk mass matrix ${\cal M}^2_{S ij}$ and no boundary masses. Again,
${\cal M}^2_{Sij}$ can be described by two mass-square eigenvalues
$M_{Si}^2$ ($i=1,2$) and a mixing angle $\theta_S$. Then the
$N$-function of $\{\varphi^i\}$ is given by \bea \label{two scalars}
N^{\lbrace \varphi^1,\varphi^2 \rbrace}(q) &=& \left(c^2
N^{\varphi(M_{S1})}_{z_1z'_1} + s^2
N^{\varphi(M_{S2})}_{z_1z'_1}\right)
\left(s^2 N^{\varphi(M_{S1})}_{z_2z'_2} + c^2 N^{\varphi(M_{S2})}_{z_2z'_2}\right) \nonumber\\
&-& c^2s^2\left( N^{\varphi(M_{S1})}_{z_1z'_2} -
N^{\varphi(M_{S2})}_{z_1z'_2}\right) \left( N^{\varphi(M_{S1})}_{z_2z'_1}
- N^{\varphi(M_{S2})}_{z_2z'_1}\right), \eea where $c=\cos\theta_S,
s=\sin\theta_S$ and $N^{\varphi(M)}_{zz^\prime}$ is the $N$ function
of a 5D scalar with orbifold parities $z,z^\prime$, which has a bulk
mass $M$ and vanishing boundary masses.

In Appendix A, we provide explicit expression of
$N^\Phi_{zz^\prime}$ for $\Phi$ with various 4D spin  and orbifold
parities, as well as its limiting behaviors at $|q|\rightarrow 0,
\infty$.
  Once
the $N$-functions are obtained, one can examine the behavior at
$q\rightarrow 0$
to find $N_0^{\{\Phi\}}$, and finally  apply  (\ref{boson_th}) to
obtain the 1-loop corrections $\Delta_a$. Using the properties of
$N$-functions described  above and also in  Appendix A, we find the
expressions of $\Delta_a^{\{A\}}$ and $\Delta_a^{\{B\}}$ presented
in Table 1 and Table 2, respectively. (See (\ref{form}) for the
definition of $\Delta_a^{\{A\},\{B\}}$.)  For the 1-loop corrections
$\Delta_a^{\{\varphi\},\{\psi\}}$   induced by scalar and fermion
fields, we consider the two cases: one for the case that there is no
mixing between matter fields with different orbifold parities, and
the other case with two scalars or two Dirac fermions which can have
such a mixing. For the first case, one can simply consider a single
scalar or a single fermion with definite orbifold parities, and the
results are summarized in Table 3 and Table 4. For the second case,
one can use  the $N$-functions (\ref{two fermions}) and
 (\ref{two scalars}) to obtain the results presented in Table 5 and 6. A prescription
for $\Delta_a^{\{\varphi\},\{\psi\}}$  in more general case is
described in Appendix A.

\section{Conclusion}
Models with warped extra dimension might provide an explanation for
various puzzles in particle physics, e.g. the weak scale to Planck
scale hierarchy  and the Yukawa coupling hierarchy, while
implementing a breaking of unified gauge symmetry in bulk spacetime
by boundary conditions, which would solve some of the naturalness
problems in grand unified theories such as the doublet-triplet
splitting problem. Kaluza-Klein threshold corrections in such models
are generically enhanced by the logarithm of an exponentially small
warp factor, and therefore can be crucial for  successful gauge
coupling unification in the framework of warped unified model. In
this paper, we discuss a novel method to compute 1-loop gauge
couplings in generic 5D gauge theory on a slice of $AdS_5$, in which
some of the bulk gauge symmetries are broken by orbifold boundary
conditions and/or by bulk Higgs vacuum values, and also there can be
nonzero mass mixings between the bulk fields with different orbifold
parities. Explicit analytic expressions of the Kaluza-Klein
thresholds as a function of various model parameters are derived,
and our analysis can cover most of the warped GUT models which have
been discussed so far in the literatures.

\ack KC and CSS are supported by the KRF grants funded by the Korean
Government (KRF-2007-341-C00010 and KRF-2008-314-C00064) and KOSEF
grant funded by the Korean Government (No. 2009-0080844). IWK is
supported by the U.S. Department of Energy under grant
 No. DE-FG02-95ER40896.

\appendix


\section{$N$ function}

\vskip 0.3cm \noindent
 In this appendix, we discuss the $N$ function of a 5D field $\Phi$
on a slice of AdS$_5$, which has a definite value of 4D spin and 4D
chirality as well as definite orbifold parities:
 \bea
 \Phi(x,y)=\{\phi,e^{-2kR|y|}\psi_L, e^{-2kR|y|}\psi_R,
 A_\mu\}
 \quad (\psi_{L,R}=\frac{1}{2}(1\pm\gamma_5)\psi ),\nonumber\eea
where the 4D scalar $\phi$ might be a 5D scalar, or the 5-th
component of a 5D vector, or a ghost field.
 Generic 5D field $\Phi$ on a slice of AdS$_5$  can be decomposed as
 \bea
\Phi(x,y)=\sum \Phi_n(x) f_n(y), \eea where the KK wavefunction
$f_n$ satisfies  \bea \left[ - e^{skR|y|}\partial_y \left(e^{-s k R
|y|}\partial_y\right) + R^2 M_\Phi^2\right]f_n = R^2 e^{2k R|y|}
m_n^2 f_n \eea  for the KK mass eigenvalue $m_n$. Here \bea
&&M_\Phi^2=\left\{{M_S^2},{M_F}\left({M_F}+k\right),{M_F}\left({M_F}-k\right),
{M_V^2}\right\},\nonumber \\
&& s=\{4,1,1,2\}\quad {\rm for}\quad \Phi=\{\phi,e^{-2kR|y|}\psi_L,
e^{-2kR|y|}\psi_R,
 A_\mu\},\eea
where $M_S,M_F$ and $M_V$ denote the bulk masses of
 $\phi,\psi$ and $A_\mu$, respectively.
Here we are using the mass parameter convention defined in
(\ref{convention1}) and (\ref{convention2}), e.g. $M_S=0$ for
$\phi=A_5^\sigma$ or $c_A^\sigma$,
 $M_S=g_{5\alpha}\lambda_\alpha$ for $\phi=c_B^\alpha$,
 $M_V=0$ for $A_\mu=A_\mu^\sigma$, and
 $M_V=g_{5\alpha}\lambda_\alpha$ for $A_\mu=B_\mu^\alpha$.

 Generic solution of the above KK equation is given by \bea f_n(y) =
e^{s k R|y|/2} \left[ A_\alpha(m_n) J_\alpha \left(\frac{m_n}{k}
e^{k R |y|}\right) + B_\alpha(m_n) Y_\alpha \left(\frac{m_n}{k} e^{k
R |y|}\right)\right], \eea where $\alpha = \sqrt{(s/2)^2 +
M_\Phi^2/k^2 }$, and  $A_\alpha, B_\alpha$ are determined by the
boundary conditions at $y=0,\pi$. To utilize those boundary
conditions, it is convenient to introduce the following functions:
\bea f_{J_0-}(q) = J_\alpha\left(\frac{q}{k}\right), &&\quad
f_{J_0+}(q) = \left[\left(r_0
-\frac{s}{2}\right)J_{\alpha}\left(\frac{q}{k}\right)
- \frac{q}{k} J'_\alpha\left(\frac{q}{k}\right)\right],\nonumber\\
f_{Y_0-}(q) = Y_\alpha\left(\frac{q}{k}\right), &&\quad f_{Y_0+}(q)
= \left[\left(r_0 -
\frac{s}{2}\right)Y_{\alpha}\left(\frac{q}{k}\right)
- \frac{q}{k} Y'_\alpha\left(\frac{q}{k}\right)\right],\nonumber\\
f_{J_\pi -}(q) = J_\alpha\left(\frac{q}{T}\right), &&\quad f_{J_\pi
+}(q) =
\left[\left(\frac{s}{2}-r_\pi\right)J_{\alpha}\left(\frac{q}{T}\right)
+ \frac{q}{T} J'_\alpha\left(\frac{q}{T}\right)\right], \nonumber\\
f_{Y_\pi -}(q) = Y_\alpha\left(\frac{q}{T}\right), &&\quad f_{Y_\pi
+}(q) =
\left[\left(\frac{s}{2}-r_\pi\right)Y_{\alpha}\left(\frac{q}{T}\right)
+ \frac{q}{T} Y'_\alpha\left(\frac{q}{T}\right)\right], \eea where
 \bea
\hskip -1.5cm T = k e^{-\pi k R},\quad
  r_0k  = \left\{{m_S},
-{M_F}, {M_F}, {m_V}\right\}, \quad r_\pi k = \left\{{\tilde{m}_S},
-{M_F}, {M_F}, {\tilde{m}_V}\right\}\eea for the boundary masses
$m_S, \tilde{m}_S$ of $\phi$ at $y=0,\pi$ and the boundary masses
$m_V,\tilde{m}_V$ of $A_\mu$  at $y=0,\pi$. Again, we are using the
mass parameter convention defined in (\ref{convention1}) and
(\ref{convention2}). Explicitly,  $m_S=m$ and
$\tilde{m}_S=\tilde{m}$ for $\phi=\varphi$, $m_S=\tilde{m}_S=2k$ for
$\phi=A_5^\sigma$, $m_S=\tilde{m}_S=0$ for $\phi=c_A^\sigma,
c_B^\alpha$,  and the boundary masses of vector field are defined as
\bea &&-\int
d^4xdy\sqrt{-G}\left(\frac{m_V}{g_5^2}\frac{\delta(y)}{\sqrt{G_{55}}}-\frac{\tilde{m}_V}{g_5^2}\frac{\delta(y-\pi)}{\sqrt{G_{55}}}\right)
G^{MN}A_M A_N, \eea which gives \bea -\int d^4x dy
\frac{e^{-2kR|y|}}{R}\left(m_V\delta(y)-\tilde
m_V\delta(y-\pi)\right)\eta^{\mu\nu}A_\mu A_\nu \eea after the field
redefinition (\ref{redef}).

Imposing the orbifold parity conditions $\Phi(-y)=z\Phi(y)$ and
$\Phi(-y+\pi)=z^\prime\Phi(y+\pi)$  gives rise to the
constraint:\bea \left(\begin{array}{cc} f_{J_0 z}(m_n) & f_{Y_0
z}(m_n)
\\  f_{J_\pi {z'}}(m_n) & f_{Y_\pi {z'}}(m_n) \end{array} \right)
\left(\begin{array}{c}  A_\alpha \\ B_\alpha \end{array}\right) = 0.
\eea  This constraint can be used to determine the KK spectrum
$\{m_n\}$, yielding \bea f_{J_0 z}(m_n)f_{Y_\pi z'}(m_n) - f_{Y_0
z}(m_n) f_{J_\pi z'}(m_n)=0. \eea This then implies that the KK
spectrum corresponds to the zeros of \bea \label{N_single}
N^\Phi_{zz'} =\pi k^{z/2} T^{z'/2} \left(f_{J_0 z}(q)f_{Y_\pi z'}(q)
- f_{Y_0 z}(q) f_{J_\pi z'}(q)\right),\eea where the prefactor $\pi
k^{z/2} T^{z'/2}$  is introduced to achieve the asymptotic behavior
\bea \frac{1}{2}\ln N^\Phi_{zz'}(i|q|)=iB_\Phi|q| +A_\Phi\ln |q|
+{\cal O}\left(|q|^{-1}\right)\quad {\rm at}\quad |q|\rightarrow
\infty.\nonumber\eea
One can confirm  that $N_{zz'}^\Phi(q)$ is a holomorphic even
function on the complex plane of $q$.

For the computation of 1-loop gauge couplings, we do not need the
full expression of the $N$-function, but the asymptotic behaviors in
the limits $|q|\rightarrow 0,\infty$. It is straightforward to find
that \bea N^\Phi_{zz'}(q) = 2q^{\frac{z+z'}{2}} \cos
\left(\frac{q(e^{\pi k R}-1)}{k} + \frac{z+z'}{4}\pi  \right) +
{\cal O}(q^{-2})\quad {\rm at}\quad |q|\rightarrow \infty, \nonumber
\eea from which we find  \bea A_\Phi=\frac{z+z^\prime}{4},\quad i
B_\Phi=\frac{e^{\pi kR}-1}{2 k}\quad {\rm for}\quad
\Phi=\{\phi,\psi_L,\psi_R,A_\mu\}.\nonumber \eea Note that the
asymptotic form of $N_{zz^\prime}^\Phi$ at $|q|\rightarrow \infty$
is independent of $\alpha, r_0$ and $r_\pi$, and therefore
independent of the bulk and boundary masses of the gauge and matter
fields in the model. It is determined just by the orbifold parities
of $\Phi$ and the background geometry, i.e. $k$ and $R$, which
affects the KK spectral density at $m_n\rightarrow \infty$. Note
also that \bea
N^\psi_{zz^\prime}=N^{\psi_L}_{zz^\prime}N^{\psi_R}_{\bar{z}\bar{z}^\prime}\quad
(\bar{z}=-z,\bar{z}^\prime=-z^\prime),\nonumber \eea and thus \bea
A_\psi=A_{\psi_L}+A_{\psi_R}=0,\nonumber \eea which means that 5D
Dirac fermion does not give rise to a logarithmic divergence.

As we have noticed,  most of the model-parameter dependence of
1-loop gauge couplings is determined by the behavior of
$N^\Phi_{zz^\prime}$ in the limit $|q|\rightarrow 0$. Specifically
it is determined by  \bea
N^\Phi_{zz^\prime}=(-q^2)^{n_0^\Phi}\left(N_0^\Phi+{\cal
O}(q^2/m_{KK}^2)\right) \quad {\rm for} \quad |q|\ll m_{KK}, \eea
where $n_0^\Phi$ is the number of zero mode from $\Phi$, and
$m_{KK}$ is the lightest KK mass. In our convention, $n_0^\Phi=0$ or
$1$. For an explicit expression of $N_0^\Phi$, let us introduce \bea
Q^{(u)}_{zz'}(x) = \frac{1}{x} \left(E^{(u)}_{zz'}(x) -
E^{(u)}_{zz'}(-x)\right), \eea where \bea
E^{(u)}_{zz'}(\alpha)=k^{z/2} T^{z'/2} e^{\alpha \pi k
R}\left(\alpha -\frac{s+u}{2} + r_0\right)^{\frac{z+1}{2}}
\left(\alpha +\frac{s+u}{2} - r_\pi\right)^{\frac{z'+1}{2}}
\nonumber \eea with the convention that $(\alpha - \frac{s+u}{2} +
r_0)^{(z+1)/2} = 1$ for $\alpha - \frac{s+u}{2} + r_0 =0, z+1=0$ and
also $(\alpha + \frac{s+u}{2} -r_\pi)^{(z^\prime+1)/2} = 1$ for
$\alpha +\frac{s+u}{2} - r_\pi =0, z^\prime +1=0$. We then find \bea
\label{N_zero} N^\Phi_{zz'}(q)= Q^\Phi_{zz'} -R^\Phi_{zz'}q^2 +
{\cal O}(q^4/m_{KK}^2) \quad {\rm at} \quad |q|\rightarrow 0,\eea
where
 \bea
Q^\Phi_{zz'} = Q^{(0)}_{zz'}(\alpha), \nonumber \\
R^\Phi_{zz'} = \frac{e^{\pi k R}}{4\alpha k^2}
\left(Q^{(2)}_{zz'}(\alpha+1) - Q^{(2)}_{zz'}(\alpha-1)\right).
 \eea

With the above results, one immediately finds that $\Phi$ does not
have a zero mode in case with $Q^\Phi_{zz^\prime}\neq 0$, and then
\bea N_0^\Phi=Q^\Phi_{zz^\prime}.\eea On the other hand, in other
case with $Q^\Phi_{zz^\prime}= 0$, there is a zero mode, and \bea
N_0^\Phi=R^\Phi_{zz^\prime}.\eea Let us derive an explicit form of
$Q^\Phi_{zz^\prime}$ and $R^\Phi_{zz^\prime}$ in some simple cases.
For $A_{\mu ++}$ with $M_V=m_V=\tilde{m}_V=0$,  $\psi_{L}$ with
$M_F\neq 0$, and $\phi$ with $M_S\neq 0$ and $m_S=\tilde{m}_S=0$, we
find \bea && Q^{A_\mu}_{++}=0,\quad N^{A_{\mu
++}}_0=R^{A_\mu}_{++}=2\pi
R e^{\pi k R/2},\nonumber \\
&&Q^{\psi_L}_{++}=0, \quad N_0^{\psi_{L++}}=R^{\psi_L}_{++}=2e^{\pi
k R/2}\left(\frac{\sinh(M_F-k/2)\pi R}{M_F-k/2}\right),\nonumber \\
&& N_0^{\psi_{L+-}}=Q^{\psi_L}_{+-}=2e^{-M_F\pi R}, \quad
N_0^{\psi_{L-+}}=Q^{\psi_L}_{-+}=2e^{M_F\pi R},\nonumber \\
&& N_0^{\phi_{++}}=Q^\phi_{++}=\frac{2M_S^2e^{-\pi
kR/2    }}{\alpha_S k}\sinh \left(\alpha_S\pi k R\right),\nonumber \\
&&N_0^{\phi_{\pm\mp}}=Q^\phi_{\pm\mp}=\frac{2e^{\pm \pi
kR/2}}{\alpha_S}\Big[\alpha_S\cosh \left(\alpha_S\pi kR\right) \mp
2\sinh \left(\alpha_S \pi k R\right)\Big],\nonumber
\\
&&N_0^{\phi_{--}}=Q^\phi_{--}=\frac{2e^{\pi kR/2}}{\alpha_S k}\sinh
\left(\alpha_S\pi k R\right),\nonumber \eea where
$\alpha_S=\sqrt{4+M_S^2/k^2}$.


  Let us now consider the $N$-function in more general case that
  there is a mass mixing between $\Phi$'s with {\it different} orbifold parities.
In such case,  the $N$-function  takes a more complicate form as the
mass eigenstate does not have a definite orbifold parity. Let
$\{\Phi_I\}$ ($I=1,2,..., n_\Phi$) denote a set of 5D fields with
the same 4D spin and unbroken gauge charges in the orbifold parity
eigenbasis, and $\{\Phi_A\}$ ($A=1,2,..., n_\Phi$) denote the same
set of fields, but in the bulk mass eigenbasis which is related to
the parity eigenbasis by a unitary rotation: \bea \Phi_A=
\sum_IU_{AI}\Phi_I.\eea Here each fermionic $\Phi_I$ is either a
left-handed spinor ($\psi_L$) or a right-handed spinor ($\psi_R$).
The KK wavefunction in the decomposition \bea \Phi_A(x,y)=\sum_n
\Phi_{An}(x)f_{An}(y)\nonumber \eea satisfies \bea \left[ -
e^{skR|y|}\partial_y \left(e^{-s k R |y|}\partial_y\right) + R^2
M_A^2\right]f_{An} = R^2 e^{2k R|y|} m_{n}^2 f_{An}, \eea  where
again \bea s=\{4,1,1,2\},\nonumber \\
M_A^2=\left\{{M_{SA}^2},M_{FA}\left({M_{FA}}+k\right),{M_{FA}}\left({M_{FA}}-k\right),
{M_{VA}^2}\right\} \eea for $\Phi_A=\{\phi_A,e^{-2kR|y|}\psi_{AL},
e^{-2kR|y|}\psi_{AR},
 A^A_\mu\}$.
 As the orbifold boundary conditions are defined in the basis
 $\{\Phi_I\}$, it is more nontrivial to find the resulting
 constraints  on the KK spectrum and the corresponding $N$-functions.
It turns out that the $N$-function in the presence of mass mixing
can be constructed with the following functions:
 \bea f^{IA}_{J_0 z_I}(q),\, f^{IA}_{Y_0 z_I}(q), \, f^{IA}_{J_\pi z'_I
}(q), \, f^{IA}_{Y_\pi z'_I}(q),\eea where $z_I,z_I^\prime$ are the
orbifold parities of $\Phi_I$ and \bea f^{IA}_{J_0-}(q) =
J_{\alpha_A}\left(\frac{q}{k}\right), &&\quad f^{IA}_{J_0+}(q) =
\left[\left(r_{0IA}-\frac{s}{2}\right)J_{\alpha_A}\left(\frac{q}{k}\right)
- \frac{q}{k} J'_{\alpha_A}\left(\frac{q}{k}\right)\right]\nonumber\\
f^{IA}_{Y_0 -}(q) = Y_{\alpha_A}\left(\frac{q}{k}\right), &&\quad
f^{IA}_{Y_0 +}(q) =
\left[\left(r_{0IA}-\frac{s}{2}\right)Y_{\alpha_A}\left(\frac{q}{k}\right)
- \frac{q}{k} Y'_{\alpha_A}\left(\frac{q}{k}\right)\right]\nonumber\\
f^{IA}_{J_\pi -}(q) = J_{\alpha_A}\left(\frac{q}{T}\right), &&\quad
f^{IA}_{J_\pi+}(q) = \left[\left(\frac{s}{2}-r_{\pi
IA}\right)J_{\alpha_A}\left(\frac{q}{T}\right)
+ \frac{q}{T} J'_{\alpha_A}\left(\frac{q}{T}\right)\right] \nonumber\\
f^{IA}_{Y_\pi -}(q) = Y_{\alpha_A}\left(\frac{q}{T}\right), &&\quad
f^{IA}_{Y_\pi +}(q) = \left[\left(\frac{s}{2}-r_{\pi
IA}\right)Y_{\alpha_A}\left(\frac{q}{T}\right) + \frac{q}{T}
Y'_{\alpha_A}\left(\frac{q}{T}\right)\right]\nonumber.  \eea Here
\bea \alpha_A = \sqrt{(s/2)^2 + M_A^2/k^2},\nonumber \eea and \bea k
r_{0IA} = \sum_J \frac{(m_{S,V})_{IJ}U^*_{AJ}}{U^*_{AI}}, \quad
 k r_{\pi IA} = \sum_J \frac{(\tilde{m}_{S,V})_{IJ}U^*_{AJ}}{U^*_{AI}}
\quad{\rm for}\quad \Phi_I=\phi_I, A_\mu^I,\nonumber\\
  k r_{0IA} = \mp M_{FA} , \quad k r_{\pi IA} = \mp M_{FA}
\quad{\rm for}\quad \Phi_I=\psi_{IL},\psi_{IR},\eea where
$(m_{S,V})_{IJ}$ and $(\tilde{m}_{S,V})_{IJ}$ are
 the boundary mass matrices  of $\phi_I, A^I_\mu$ at $y=0$ and $y=\pi$, respectively, defined in
the orbifold parity eigenbasis.

Then, for the KK wavefunction \bea f_{An}(y) = e^{s k R|y|/2} \left[
A_{A}(m_n) J_\alpha \left(\frac{m_n}{k} e^{k R |y|}\right) +
B_{A}(m_n) Y_\alpha \left(\frac{m_n}{k} e^{k R
|y|}\right)\right],\nonumber  \eea the orbifold boundary conditions
yield
\bea  \label{obc} {\cal B}\left(\begin{array}{c} A_{A} \\
B_{A}
\end{array}\right) =\left(\begin{array}{cc}B_{J_0} & B_{Y_0} \\
B_{J_\pi} & B_{Y_\pi}
\end{array}\right)\left(\begin{array}{c} A_A \\B_A \end{array}\right)
=0,
\eea where ${\cal B}$ is a $2n_\Phi\times 2n_\Phi$ matrix given by
\bea \left(B_{J_0}\right)_{IA} = U^*_{AI} f^{IA}_{J_0 z_I}, \quad
\left(B_{Y_0}\right)_{IA} = U^*_{AI} f^{ IA}_{Y_0 z_I}, \nonumber \\
\left(B_{J_\pi}\right)_{IA} = U^*_{AI} f^{IA}_{J_\pi z'_I}, \quad
\left(B_{Y_\pi}\right)_{IA} = U^*_{AI} f^{IA}_{Y_\pi z'_I} \quad
\textrm{for} \,\,\, \Phi_I=\phi_I\,\,\,{\rm or}\,\,\,
A_\mu^I,\nonumber \eea and \bea \left(B_{J_0}\right)_{IA} =
(UF^{L,R}_0)^*_{AI} f^{IA}_{J_0 z_I}, \quad
  \left(B_{Y_0}\right)_{IA} = (UF^{L,R}_0)^*_{AI} f^{IA}_{Y_0 z_I},\nonumber \\
\left(B_{J_\pi}\right)_{IA} = (UF^{L,R}_\pi)^*_{AI}
f^{IA}_{J_\pi z'_I}, \quad \left(B_{Y_\pi}\right)_{IA} =
(UF^{L,R}_\pi)^*_{AI}
f^{IA}_{Y_\pi z'_I } \,\,\,{\rm for}\,\,\, \Phi_I=\psi_{IL},\psi_{IR},\nonumber\eea
where
 \bea
\left(F^{L,R}_{0,\pi}\right)_{IJ}=\frac{\left({\cal
F}^{L,R}_{0,\pi}\right)_{IJ}}{[{\rm det}({\cal
F}^{L,R}_{0,\pi})]^{1/n_\Phi}},\nonumber \\
\left({\cal F}^{L,R}_{0}\right)_{IJ}=\delta_{IJ}\pm
z_I{\mu}_{IJ},\quad \left({\cal
F}^{L,R}_{\pi}\right)_{IJ}=\delta_{IJ}\pm z^\prime_I\tilde{\mu}_{IJ}
\nonumber \eea
 for the boundary fermion
masses $\mu_{IJ},\tilde{\mu}_{IJ}$ defined in (\ref{convention2}).

With (\ref{obc}), the $N$ function of $\{\Phi_I\}$ is proportional
to the determinant of the $2n_\Phi\times 2n_\Phi$ matrix ${\cal B}$.
In fact, one can show that the $N$-function can be reduced to the
determinant of an $n_\Phi\times n_\Phi$ matrix: \bea
N^{\{\Phi\}}(q)={\rm det}\left(B_N(q)\right)\eea where \bea
\label{bn} \left(B_N\right)_{IJ}=\sum_A
U^{*}_{AI}U_{AJ}N^{IJ,A}_{z_Iz^\prime_J}(q)\quad {\rm for}\quad
\Phi_I=\phi_I\,\,\, {\rm or}\,\,\, A^I_\mu,\nonumber \\
\left(B_N\right)_{IJ}=\sum_A
(UF^{L,R}_0)^*_{AI}(UF^{L,R}_\pi)_{AJ}N^{IJ,A}_{z_Iz^\prime_J}(q)\quad
{\rm for}\quad \Phi_I=\psi_{IL},\psi_{IR} \eea with \bea
N^{IJ,A}_{z_Iz'_J}(q) = \pi  k^{z_I/2}T^{z_J'/2} \left(f^{IA}_{J_0
z_I}(q)f^{JA*}_{Y_\pi z'_J}(q) - f^{IA}_{Y_0 z_I}(q) f^{JA *}_{J_\pi
z'_J}(q)\right). \eea Note that this function is nothing but the $N$
function defined in (\ref{N_single}) with $\alpha\rightarrow
\alpha_A$, $r_0\rightarrow r_{0IA}$, $r_\pi\rightarrow r_{\pi IA}$
and $z,z^\prime\rightarrow z_I,z_I^\prime$. Furthermore its limiting
behavior at $|q|\rightarrow \infty$ is independent of $\alpha_A,
r_{0IA}, r_{\pi IA}$: \bea \frac{1}{2}\ln
N^{IJ,A}_{z_Iz^\prime_J}(i|q|)=\frac{e^{\pi kR}-1}{2
k}|q|+\frac{z_I+z_J^\prime}{4}\ln|q|+{\cal O}(|q|^{-1}),
\nonumber \eea and therefore  \bea \frac{1}{2}\ln
N^{\{\Phi\}}(i|q|)=\frac{n_\Phi (e^{\pi kR}-1)}{2
k}|q|+\sum_I\frac{(z_I+z_I^\prime)}{4}\ln|q|+{\cal
O}(|q|^{-1})\nonumber\eea at $|q|\rightarrow \infty$.

 To obtain the
1-loop corrections to low energy gauge couplings induced by
$\{\Phi\}$, we need to know the limiting behavior of $N^{\{\Phi\}}$
at $q\rightarrow 0$: \bea N^{\{\Phi\}}={\rm
det}(B_N)=(-q^2)^{n_0^{\{\Phi\}}}\left(N_0^{\{\Phi\}}+{\cal
O}(q^2/m_{KK}^2)\right).\nonumber \eea It is straightforward to find
$N_0^{\{\Phi\}}$ from the limiting behavior  of $N^\Phi_{zz^\prime}$
in (\ref{N_zero}) and the expression  of $(B_N)_{IJ}$ in (\ref{bn}).

%

\section{KK thresholds with boundary matter fields}

In this paper, we did not include a boundary matter field
separately. In fact, boundary matter field can always be considered
as a 4D mode of bulk matter field localized at the boundary in the
limit that the 5D mass of bulk field approaches to the cutoff scale
$\Lambda$\footnote{For scalar field, we also need proper boundary
masses comparable to $\Lambda$.}. This means that the 1-loop gauge
coupling in the presence of boundary matter field can be obtained
from our results by taking an appropriate limit. Here we discuss
this point with simple examples in flat spacetime background.

Let us first consider a Dirac fermion $\psi_{++}$ with bulk mass
$M_F$. By taking the limit $k\rightarrow 0$ for the result in Table
4, one easily finds that the 1-loop correction due to $\psi_{++}$ is
given by \bea \frac{1}{8\pi^2}\Delta_a^{\psi_{++}}=
\frac{1}{12\pi^2}{\rm
Tr}(T_a^2(\psi))\left[\ln\frac{M_F}{p}-\ln\left(\sinh M_F\pi
R\right)\right].\eea In the limit
 $M_F\rightarrow \Lambda\gg 1/R$, the chiral zero mode becomes localized at $y=0$.
 On the other hand, all KK modes get a mass comparable to $\Lambda$,
 and therefore can be integrated out while leaving a trace only in
  the Wilsonian couplings at $\Lambda$.
 Indeed
$\Delta_a^{\psi_{++}}$ in the limit $M_F\rightarrow \Lambda$ becomes
the 1-loop correction due to a 4D boundary chiral fermion after
subtracting the power-law divergence which should be absorbed into
the renormalization of the 5D gauge coupling $1/g_{5a}^2$ at
$\Lambda$: \bea \frac{1}{8\pi^2}\Delta_a^{\psi_{++}}\rightarrow
\frac{1}{12\pi^2}{\rm
Tr}(T_a^2(\psi))\left[\ln\frac{\Lambda}{p}-\Lambda \pi R+{\cal
O}(1)\right]. \eea

As another example, let us consider $\psi_{+-}$ with bulk mass
$M_F$, which gives a correction \bea
\frac{1}{8\pi^2}\Delta_a^{\psi_{+-}}=\frac{1}{12\pi^2}{\rm
Tr}(T_a^2(\psi))M_F\pi R.\eea  In the limit $M_F \rightarrow
\Lambda\gg 1/R$, there appear two chiral fermion modes localized at
the boundaries, one at $y=0$ and another at $y=\pi$, which form a 4D
Dirac fermion with 4D mass $m_D=2M_F e^{-M_F\pi R}$, while all other
modes have a mass of ${\cal O}(\Lambda)$. As the above 1-loop gauge
coupling assumes that there is no light mode with a mass lighter
than the external momentum $p$ of the gauge boson zero mode, it can
be directly used only for $p<m_D$. We then find \bea
\frac{1}{8\pi^2}\Delta_a^{\psi_{+-}}\rightarrow
\frac{1}{12\pi^2}{\rm Tr}(T_a^2(\psi))\left[
2\ln\frac{\Lambda}{m_D}-\Lambda\pi R +{\cal O}(1)\right],\eea which
corresponds, after subtracting  the power law divergence, to the
1-loop threshold due to a massive 4D Dirac fermion with mass $m_D$.
 If we consider the limit that $m_D$ becomes even smaller than $p$, the IR cutoff of the momentum integral of the localized modes
should be taken as $p$, and then we arrive at the standard 1-loop
correction due to a massless 4D Dirac fermion: \bea
\frac{1}{8\pi^2}\Delta_a= \frac{1}{6\pi^2}{\rm
Tr}(T_a^2(\psi))\ln(\Lambda/p).\eea

Let us finally consider the case of two 5D Dirac fermions
$\psi^1_{++}$ and $\psi^2_{--}$ with a diagonal 5D mass matrix
$M_{Fpq}=M_{Fp}\delta_{pq}$ ($p,q=1,2$) and a boundary mass-mixing
\bea \int d^4x dy \delta(y)2\mu
\left(\bar{\psi}^1\psi^2+\bar{\psi}^2\psi^1\right).\eea (Note that
5D fermion has a mass-dimension 2, and thus $\mu$ is a dimensionless
parameter in our convention.) With the results in Table 6, one
easily finds \bea \frac{1}{8\pi^2}\Delta_a^{\{\psi^1,\psi^2\}}=
-\frac{1}{12\pi^2} {\rm
Tr}(T_a^2(\psi))\ln\left[\left(\frac{\mu^2}{1+\mu^2}\right)e^{(M_{F1}-M_{F2})\pi
R}\right].\eea In the limit $M_{F2}\rightarrow -\Lambda$ with
$\mu\ll 1$, $\psi_{--}^2$ gives a chiral zero mode $\chi$ localized
at $y=0$, while all other modes of $\psi_{--}$ are decoupled with a
mass comparable to $\Lambda$. The resulting effective theory
contains a 5D fermion $\psi^1_{++}$ and a chiral boundary fermion
$\chi$ with a mass mixing:\bea \int d^4x dy \delta(y)2\mu_{\rm eff}
\left(\bar{\psi}^1\chi+\bar{\chi}\psi^1\right),\eea where $\mu_{\rm
eff}=2\mu\sqrt{\Lambda}\ll \sqrt{\Lambda}$ for the canonically
normalized 4D fermion $\chi$. In the same limit, \bea
\frac{1}{8\pi^2}\Delta_a^{\{\psi^1,\psi^2\}}\rightarrow
\frac{1}{12\pi^2} {\rm
Tr}(T_a^2(\psi))\left[\ln\left(\frac{\Lambda}{\mu_{\rm
eff}^2}\right) - \ln \left(e^{M_{F1}\pi R}\right)-\Lambda\pi R
+{\cal O}(1)\right], \nonumber \eea  which corresponds to the 1-loop
threshold in the effective theory again after subtracting the power
law divergence.

\pagebreak

\begin{table}
{\bf Table 1.} 1-loop gauge couplings induced by 5D vector fields
$A_M^\sigma$ in the limit $p \ll m_{KK}$ where $m_{KK}$ is the
lowest nonzero KK mass. Here $c_a(A)={\rm Tr}(T_a^2(A_M))$.
\bigskip

\begin{center}
\begin{tabular}{|c|l|}
\hline $\begin{array}{c} \\ {} \end{array} (zz')
\begin{array}{c} \\ {} \end{array}$ & \hskip 4cm $\Delta_a^{\{A\}}$ \\
\hline \hline $\begin{array}{c} \\ {} \end{array} (++)
\begin{array}{c} \\ {} \end{array}$
&$\frac{c_a(A)}{12} \left[ 22  \pi k R -23 \ln
\left(\Lambda \pi R\right) + 44\ln (p\pi R)\right]  $ \\

\cline{1-2}

$\begin{array}{c} \\ {} \end{array} (+-)
\begin{array}{c} \\ {} \end{array}$
& $\frac{c_a(A)}{12} \left( - 22\pi k R \right)$ \\

\cline{1-2}

$\begin{array}{c} \\ {} \end{array} (-+)
\begin{array}{c} \\ {} \end{array}$
& $\frac{c_a(A)}{12} \left( 22\pi k R\right) $\\

\cline{1-2}

$\begin{array}{c} \\ {} \end{array} (--)
\begin{array}{c} \\ {} \end{array}$
& $\frac{c_a(A)}{12} \left[ 21 \ln\left(\frac{\sinh\pi k R }{\pi k
R}\right)- \pi k R + 23 \ln
\left(\Lambda \pi R\right) -2\ln(p\pi R)\right]$\\

\hline

\end{tabular} \vskip 2.5cm
\end{center}
{\bf Table 2.} 1-loop gauge couplings induced by 5D vector fields
$B_M^\alpha$ and Goldstone bosons $\pi^\alpha$ for the range of
$M_B$ which does not give any zero mode lighter than $p$. Here
$c_a(B)={\rm Tr}(T_a^2(B_M))$, $M_B = g_{5\alpha}\lambda_\alpha$,
and $\alpha_B = \sqrt{1+ M_{B}^2/k^2}$, where  $M_B$ is the
canonical 5D mass of $B^\alpha_M$.
\bigskip

\begin{center}
\begin{tabular}{|c|l|}
\hline $\begin{array}{c} \\ {} \end{array} (zz')
\begin{array}{c} \\ {} \end{array}$ & \hskip 4.2cm $\Delta^{\{B\}}_a$ \\
\hline \hline $\begin{array}{c} \\ {} \end{array} (++)
\begin{array}{c} \\ {} \end{array}$
&$\frac{c_a(B)}{12}
 \left[ 20 \ln \left(\frac{\sinh\alpha_B\pi k R }{\alpha_B  \pi k R}\right)
+ 42 \ln\left(M_{B}\pi R\right) - 22 \ln \left(\Lambda\pi R\right) \right]  $ \\

\cline{1-2}

$\begin{array}{c} \\ {} \end{array} (+-)
\begin{array}{c} \\ {} \end{array}$
& $\frac{c_a(B)}{12}
\left[20 \ln \left(\frac{\alpha_B \cosh\alpha_B\pi k R  - \sinh\alpha_B\pi k R}{ \alpha_B}\right) \right]$ \\

\cline{1-2}

$\begin{array}{c} \\ {} \end{array} (-+)
\begin{array}{c} \\ {} \end{array}$
& $\frac{c_a(B)}{12}
\left[ 20 \ln \left(\frac{\alpha_B \cosh\alpha_B\pi k R  + \sinh\alpha_B\pi k R}{ \alpha_B}\right)\right] $\\

\cline{1-2}

$\begin{array}{c} \\ {} \end{array} (--)
\begin{array}{c} \\ {} \end{array}$
& $\frac{c_a(B)}{12} \left[ 20 \ln \left(\frac{\sinh\alpha_B \pi k
R}{\alpha_B \pi k R}\right) - 2 \ln(M_{B}\pi R) + 22 \ln (\Lambda
\pi R)
\right]$\\

\hline

\end{tabular}
\end{center}
\end{table}

\begin{table}
{\bf Table 3.} 1-loop gauge couplings  induced by a 5D real scalar
$\varphi$ with definite orbifold parities $z,z^\prime$. Here
$c_a(\varphi)={\rm Tr}(T_a^2(\varphi))$, $\alpha=\sqrt{4+M_S^2/k^2}
$, where $M_S$,
$m_S$ and $\tilde m_S$ are the bulk and boundary masses  of
$\varphi$. $\varphi^{(0)}$ denotes a particular type of 5D scalar
field with $(zz^\prime)=(++)$, $m_S=\tilde{m}_S$ and
$M_S^2=m_S(m_S-4k)$, which has a zero mode lighter than $p$.

\bigskip
\begin{center}
\begin{tabular}{|c|l|}
\hline $\begin{array}{c}\\{}\end{array} (zz') \begin{array}{c}\\{}\end{array}$
& \hskip 6cm$\Delta_a^{\{\varphi\}}$ \\
\hline \hline $\begin{array}{c}{}\\{}\\ (++) \\{} \\{}\end{array}$ &
$\begin{array}{l} \frac{-1}{12}c_a(\varphi^{(0)}) \left[\ln
\left(\frac{\sinh{(m_S-k)\pi R} }{(m_S-k)\pi R
}\right) + \pi k R
-  \ln(\Lambda \pi R) + 2\ln(p\pi R) \right] \\ {}\\
\frac{-1}{12}c_a(\varphi) \left[\ln \left(\frac{\alpha k(m_S-\tilde m_S)\cosh{\alpha\pi k R}+(\alpha^2 k^2 - (2k -m_S)(2k-\tilde m_S))
\sinh{\alpha\pi k R} }{\alpha k }\right) -\ln\Lambda \right] \end{array} $ \\

\cline{1-2}

$\begin{array}{c} \\{}\end{array} (+-)\begin{array}{c}\\{}
\end{array}$ &
$\frac{-c_a(\varphi)}{12}\left[\ln\left(\frac{\alpha k\cosh\alpha\pi k R
- (2k-m_S)\sinh\alpha\pi k R}{\alpha k} \right) \right]$
 \\

\cline{1-2} $\begin{array}{c}\\{}\end{array}
(-+)\begin{array}{c}\\{} \end{array}$ &
$\frac{-c_a(\varphi)}{12}\left[\ln\left(\frac{\alpha k\cosh\alpha\pi k
R + (2k-\tilde m_S)\sinh\alpha\pi k R}{\alpha k}
\right) \right]$ \\

\cline{1-2}

$\begin{array}{c}\\{}\end{array} (--) \begin{array}{c}\\{}\end{array}$
 & $\frac{-c_a(\varphi)}{12}\left[\ln\left(\frac{\sinh\alpha\pi k
R}{\alpha \pi k R}\right)  +\ln(\Lambda\pi R)\right]$ \\
 \hline
\end{tabular}
\end{center}
\vskip 2.0cm

{\bf Table 4.} 1-loop gauge couplings induced by a Dirac fermion $\psi$
with definite orbifold parities. Here $c_a(\psi)={\rm
Tr}(T_a^2(\psi))$ and $M_F$ is the bulk
mass of $\psi$.
\bigskip
\begin{center}
\begin{tabular}{|c|l|}
\hline $\begin{array}{c}\\{}\end{array} (zz') \begin{array}{c}\\{}\end{array}$
& \hskip 3cm$\Delta_a^{\{\psi\}}$ \\
\hline \hline $\begin{array}{c}\\{}\end{array}(++) \begin{array}{c}\\{} \end{array}$ &
$-\frac{2}{3}c_a(\psi) \left[\ln \left(\frac{\sinh{(M_F-k/2)\pi  R} }{(M_F-k/2)\pi  R }\right) + \frac{1}{2}\pi k R
+  \ln(p \pi R) \right]$ \\

\cline{1-2}

$\begin{array}{c} \\{}\end{array} (+-)\begin{array}{c}\\{} \end{array}$
& $+\frac{2}{3}c_a(\psi)\ M_F \pi R $ \\

\cline{1-2}
$\begin{array}{c}\\{}\end{array} (-+)\begin{array}{c}\\{} \end{array}$
& $-\frac{2}{3}c_a(\psi) \ M_F\pi  R $ \\

\cline{1-2}

$\begin{array}{c}\\{}\end{array} (--) \begin{array}{c}\\{}\end{array}$
 & $-\frac{2}{3}c_a(\psi)
\left[\ln \left(\frac{\sinh{(M_F+k/2) \pi  R}
}{(M_F + k/2) \pi  R}\right)
+ \frac{1}{2}\pi k R +  \ln(p\pi R) \right] $ \\
 \hline
\end{tabular}
\end{center}
\end{table}

\begin{table}
{\bf Table 5.}  1-loop corrections induced by two real scalars
$\{\varphi_{z_1z'_1}^1,\varphi_{z_2z'_2}^2\}$ which have the same
gauge charge, but can have different orbifold parities. Here
$\alpha_{i}=\sqrt{4+M_{Si}^2/k^2}$ ($i=1,2$) for the bulk mass
eigenvalues  $M_{Si}$, and $s\equiv \sin\theta_S$, $c\equiv
\cos\theta_S$ for the mixing angle $\theta_S$. We are considering a
generic bulk mass matrix which does not give any zero mode lighter
than $p$, while the boundary masses are assumed to be zero for
simplicity.
\bigskip
\begin{center}
\begin{tabular}{|c|l|}
\hline $\begin{array}{c}(z_1z_1') \\ (z_2z_2')\end{array}$
& \hskip 5.5cm $\Delta_a^{\{\varphi^1,\varphi^2\}}$ \\
\hline \hline $\begin{array}{c}(++) \\ (++)\\ \end{array}$ &
$\frac{-c_a(\varphi)}{12}\left[\ln\left(\frac{\sinh\alpha_1\pi k
R}{\alpha_1 \pi k R}\right)\left(\frac{\sinh\alpha_2\pi k
R}{\alpha_2 \pi k R}\right) + 2\ln(M_{S1} \pi R)(M_{S2}\pi
R)-2\ln(\Lambda\pi R)\right]$ \\

\cline{1-2}

$\begin{array}{c} \\ (++)\end{array}$& $
\frac{-c_a(\varphi)}{12}\left[\ln\left\lbrace
c^2\left(\frac{(M_{S1}\pi R)^2\sinh\alpha_1\pi k R}{\alpha_1 \pi k R
}\right)\left(\frac{\alpha_2\cosh\alpha_2\pi k R -
2\sinh\alpha_2\pi k R}{\alpha_2 }\right) \right.\right.$ \\
 $\begin{array}{c} (+-) \\ {} \end{array}$&
  $\quad\quad\quad\quad \ \left.\left. +
s^2\left(\frac{(M_{S2}\pi R)^2\sinh\alpha_2\pi k R}{\alpha_2 \pi k R
}\right)\left(\frac{\alpha_1\cosh\alpha_1\pi k R -
2\sinh\alpha_1\pi k R}{\alpha_1 }\right) \right\rbrace
-\ln(\Lambda\pi R)\right] $
\\

\cline{1-2}

$\begin{array}{c} \\ (++)\end{array}$& $
\frac{-c_a(\varphi)}{12}\left[\ln\left\lbrace
c^2\left(\frac{(M_{S1}\pi R)^2\sinh\alpha_1\pi k R}{\alpha_1 \pi k R
}\right)\left(\frac{\alpha_2\cosh\alpha_2\pi k R +
2\sinh\alpha_2\pi k R}{\alpha_2 }\right) \right.\right.$ \\
 $\begin{array}{c} (-+) \\ {} \end{array}$&
  $\quad\quad\quad\quad \ \left.\left. +
s^2\left(\frac{(M_{S2}\pi R)^2\sinh\alpha_2\pi k R}{\alpha_2 \pi k R
}\right)\left(\frac{\alpha_1\cosh\alpha_1\pi k +
2\sinh\alpha_1\pi k R}{\alpha_1 }\right) \right\rbrace
-\ln(\Lambda\pi R)\right] $
\\

\cline{1-2}

$\begin{array}{c} (++) \\ (--)\end{array}$ & $
\frac{-c_a(\varphi)}{12}\ln\left\lbrace((\alpha_1 c^2 + \alpha_2
s^2)^2 -4)\left(\frac{\sinh\alpha_1\pi k R}{\alpha_1}\frac{
\sinh\alpha_2\pi k R}{\alpha_2}\right) +
4c^2s^2\sinh^2\frac{(\alpha_1-\alpha_2)\pi k R}{2}\right\rbrace$\\

\cline{1-2} $\begin{array}{c} (+-) \\ (+-)\end{array}$ &
$\frac{-c_a(\varphi)}{12}\left[\ln\left(\frac{\alpha_1\cosh\alpha_1\pi
k R - 2\sinh\alpha_1\pi k R}{\alpha_1 }
\right)+\ln\left(\frac{\alpha_2\cosh\alpha_2\pi k R -
2\sinh\alpha_2\pi k R}{\alpha_2 }\right) \right]$
\\

\cline{1-2}

$\begin{array}{c}  \\ {}\end{array}$ & $\frac{-c_a(\varphi)}{12}
\ln\left\lbrace\left(\frac{\alpha_1\cosh\alpha_1\pi k R -
2\sinh\alpha_1\pi k R}{\alpha_1
}\right)\left(\frac{\alpha_2\cosh\alpha_2\pi k R + 2\sinh\alpha_2\pi
k R}{\alpha_2 }\right) \right.$\\
 $\begin{array}{c} (+-) \\ (-+)\end{array}$ & $\quad\quad\quad\quad
\ +4 s^2\left(\frac{\alpha_2\sinh\alpha_1\pi k R \cosh\alpha_2\pi k
R - \alpha_1\sinh\alpha_2\pi k R\cosh\alpha_1\pi k
R}{\alpha_1\alpha_2}\right)$ \\
$\begin{array}{c}  \\
{}\end{array}$ & $\quad\quad\quad\quad \
\left.+(\alpha_1-\alpha_2)^2 c^2s^2\left(\frac{\sinh\alpha_1\pi k
R}{\alpha_1}\frac{ \sinh\alpha_2\pi k R}{\alpha_2}\right) -
4c^2s^2\sinh^2\frac{(\alpha_1-\alpha_2)\pi k R}{2}
\right\rbrace$ \\

\cline{1-2} $\begin{array}{c}  \\ (+-)\end{array}$ &
$\frac{-c_a(\varphi)}{12}\left[\ln\left\lbrace c^2 \left(
\frac{\alpha_1\cosh\alpha_1\pi k R - 2\sinh\alpha_1\pi k R}{\alpha_1
}\right)\left( \frac{\sinh\alpha_2\pi k R}{\alpha_2 \pi
k R }\right) \right.\right. $ \\
 $\begin{array}{c} (--) \\
{}\end{array}$ & $\quad\quad\quad\quad \ \left.\left. +
s^2\left(\frac{\alpha_2\cosh\alpha_2\pi k R - 2\sinh\alpha_2\pi k
R}{\alpha_2 }\right)\left( \frac{\sinh\alpha_1\pi k R}{\alpha_1 \pi
k R }\right) \right\rbrace +\ln(\Lambda\pi R)\right] $
\\

\cline{1-2} $\begin{array}{c} (-+) \\ (-+)\end{array}$ &
$\frac{-c_a(\varphi)}{12}\left[\ln\left(\frac{\alpha_1\cosh\alpha_1\pi
k R + 2\sinh\alpha_1\pi k R}{\alpha_1 }
\right)+\ln\left(\frac{\alpha_2\cosh\alpha_2\pi k R +
2\sinh\alpha_2\pi k R}{\alpha_2 }\right) \right]$
\\

\cline{1-2} $\begin{array}{c}  \\ (-+)\end{array}$ &
$\frac{-c_a(\varphi)}{12}\left[\ln\left\lbrace c^2 \left(
\frac{\alpha_1\cosh\alpha_1\pi k R + 2\sinh\alpha_1\pi k R}{\alpha_1
}\right)\left( \frac{\sinh\alpha_2\pi k R}{\alpha_2 \pi
k R }\right) \right.\right. $ \\
 $\begin{array}{c} (--) \\
{}\end{array}$ & $\quad\quad\quad\quad \ \left.\left. +
s^2\left(\frac{\alpha_2\cosh\alpha_2\pi k R + 2\sinh\alpha_2\pi k
R}{\alpha_2 }\right)\left( \frac{\sinh\alpha_1\pi k R}{\alpha_1 \pi
k R }\right) \right\rbrace +\ln(\Lambda\pi R)\right] $
\\

\cline{1-2}

$\begin{array}{c} (--) \\ (--)\end{array}$ &
$\frac{-c_a(\varphi)}{12}\left[\ln\left(\frac{\sinh\alpha_1\pi k
R}{\alpha_1 \pi k R}\right)\left(\frac{\sinh\alpha_2\pi k
R}{\alpha_2 \pi k
R}\right)  +2\ln(\Lambda\pi R)\right]$ \\
 \hline
\end{tabular}
\end{center}
\end{table}

\begin{table}
{\bf Table 6.} 1-loop corrections induced by two Dirac fermions
$\{\psi_{z_1z'_1}^1,\psi_{z_2z'_2}^2\}$ which have the same gauge
charge, but can have different orbifold parities. Here $M_{Fp}$
$(p=1,2)$ are the bulk mass eigenvalues, and $c_{0,\pi}, s_{0,\pi}$
are defined in (\ref{cs}) in
terms of the bulk mixing angle and the boundary mass mixings. We are
considering a generic parameter range in which all nonzero KK masses
are heavier than $p$.
\bigskip
\begin{center}
\begin{tabular}{|c|l|}
\hline $\begin{array}{c}(z_1z_1') \\ (z_2z_2')\end{array}$ & \hskip 5.5cm$\Delta_a^{\{\psi^1,\psi^2\}}$ \\
\hline \hline $\begin{array}{c}(++) \\ (++)\\ \end{array}$ &
$-\frac{2}{3}c_a(\psi) \left[\ln \left(\frac{\sinh{(M_{F1}-k/2)\pi  R} }
{(M_{F1}-k/2)\pi  R }\right) +
\ln \left(\frac{\sinh{(M_{F2}-k/2)\pi  R}}{(M_{F2}-k/2)\pi  R}\right)+\pi k R
+ 2 \ln(p \pi R) \right]$ \\

\cline{1-2}

$\begin{array}{c} \\ (++)\end{array}$ & $-\frac{2}{3}c_a(\psi)
\left[\ln\left\lbrace |c_\pi|^2 \left(\frac{\sinh{(M_{F1}
-k/2)\pi  R}}{(M_{F1}
-k/2)\pi  R}\right)
e^{-M_{F2} \pi R} \right.\right.
$ \\
 $\begin{array}{c} (+-) \\ {} \end{array}$&
  $ \quad\quad\quad\quad\ \ \  \left. \mathinner{+} \left. |s_\pi|^2
\left(\frac{\sinh{(M_{F2}-k/2)\pi  R}}{(M_{F2}-k/2)\pi  R}\right)
e^{- M_{F1} \pi R}\right\rbrace + \frac{1}{2}\pi k R  +  \ln(p\pi R) \right]
  $ \\

\cline{1-2}

$\begin{array}{c} \\ (++)\end{array}$ &$-\frac{2}{3}c_a(\psi)
\left[\ln\left\lbrace |c_0|^2 \left(\frac{\sinh{(M_{F1}-k/2)\pi R}}
{(M_{F1}-k/2)\pi R}\right)
e^{M_{F2} \pi  R} \right.\right.$ \\
 $\begin{array}{c} (-+) \\ {} \end{array}$&
  $ \quad\quad\quad\quad\ \ \  \left. \mathinner{+} \left.
  |s_0|^2 \left(\frac{\sinh{(M_{F2}-k/2)\pi
R}}{(M_{F2}-k/2)\pi  R}\right) e^{ M_{F1} \pi R}\right\rbrace
+ \frac{1}{2}\pi k R  +  \ln(p \pi R) \right]
   $
\\

\cline{1-2}

$\begin{array}{c} (++) \\ (--)\end{array}$ & $ -\frac{2}{3}
c_a(\psi) \ln\left|c_0 s_\pi e^{-(M_{F1}-M_{F2}) \pi  R/2} -
c_\pi s_0 e^{(M_{F1}-M_{F2}) \pi R/2}\right|^2  $
\\

\cline{1-2}

$\begin{array}{c} (+-)\\(+-) \end{array}$
& $+\frac{2}{3}c_a(\psi) ( M_{F1}\pi R + M_{F2}\pi R)$ \\

\cline{1-2}

 $\begin{array}{c} (+-) \\ (-+)\end{array}$
 & $-\frac{2}{3}c_a(\psi) \ln \left| c_0 c^*_\pi e^{-(M_{F1}-M_{F2}) \pi  R/2}
 + s_0  s^*_\pi e^{(M_{F1}-M_{F2}) \pi  R/2}\right|^2 $ \\

\cline{1-2}

$\begin{array}{c}  \\ (+-)\end{array}$ &
$-\frac{2}{3}c_a(\psi)\left[ \ln\left\lbrace |c_0|^2
\left(\frac{\sinh{(M_{F2}+k/2)\pi R}}{(M_{F2}+k/2) \pi  R}\right)e^{-M_{F1} \pi  R} \right.
\right.
$ \\
 $\begin{array}{c} (--) \\
{}\end{array}$ & $\quad\quad\quad\quad\ \ \  \left. \mathinner{+}
\left. |s_0|^2 \left(\frac{\sinh{(M_{F1}+k/2)\pi R}}{(M_{F1}+k/2) \pi  R}\right)e^{-M_{F2} \pi  R}
\right\rbrace + \frac{1}{2}\pi k
R  + \ln( p \pi  R) \right] $
\\

\cline{1-2}
$\begin{array}{c} (-+)\\(-+) \end{array}$
& $-\frac{2}{3}c_a(\psi) ( M_{F1}\pi R + M_{F2}\pi R ) $ \\

\cline{1-2} $\begin{array}{c}  \\ (-+)\end{array}$ &
$-\frac{2}{3}c_a(\psi)\left[\ln\left\lbrace |c_\pi|^2
\left(\frac{\sinh{(M_{F2}+k/2)\pi R}}{(M_{F2}+k/2) \pi R}\right)e^{M_{F1} \pi  R} \right.
\right.
$ \\
 $\begin{array}{c} (--) \\
{}\end{array}$ & $\quad\quad\quad\quad\ \ \  \left.\mathinner{+}
\left. |s_\pi|^2 \left(\frac{\sinh{(M_{F1}+k/2)\pi R}}{(M_{F1}+k/2) \pi  R}\right)e^{M_{F2} \pi  R}
\right\rbrace  + \frac{1}{2}\pi k
R  + \ln( p\pi  R) \right] $
\\

\cline{1-2}

$\begin{array}{c} (--) \\ (--)\end{array}$ & $-\frac{2}{3}c_a(\psi)
\left[\ln \left(\frac{\sinh{(M_{F1}+k/2) \pi  R}}{(M_{F1}+k/2) \pi R}\right)
+  \ln \left(\frac{\sinh{(M_{F2}+k/2) \pi  R}}{(M_{F2}+k/2)  \pi  R}\right)
+ \pi k R + 2 \ln(p\pi R) \right] $ \\
 \hline
\end{tabular}
\end{center}
\end{table}

\pagebreak

\section*{References}

\end{document}